\documentclass[sigconf]{acmart}

\acmYear{2017}

\usepackage{epsfig,endnotes}

\usepackage{amsmath, amssymb, amsfonts}
\usepackage{epsfig, graphicx, subfigure}
\usepackage{mdwlist}
\usepackage{algorithmic}
\usepackage{wrapfig}
\usepackage{algorithm,booktabs}
\usepackage{latexsym}
\usepackage{comment}
\usepackage{url}
\usepackage{subfigmat}
\usepackage{etoolbox}
\usepackage{float}
\usepackage{subfigure}

\usepackage{multirow}
\usepackage{graphicx}
\usepackage{hhline}
\usepackage{color}
\usepackage{enumitem}

\begin{document}

%\title{Study of End-to-End Encrypted Messaging Apps}
%\title{Security-Usability Pitfalls of Key-Fingerprint Verification in End-to-End Encrypted Remote Communication Apps
%%\thanks{Confidential copy -- Please do not duplicate or distribute.}
%}
\title{On the Pitfalls of End-to-End Encrypted Communications: \\A Study of Remote Key-Fingerprint Verification}

%\numberofauthors{8}

\author{Maliheh Shirvanian, Nitesh Saxena, and Jesvin James George\\
{University of Alabama at Birmingham} \\
Email: maliheh, saxena, jesvin@uab.edu}

%
%\begin{CCSXML}
%<ccs2012>
%<concept>
%<concept_id>10002978.10003006.10003013</concept_id>
%<concept_desc>Security and privacy~Distributed systems security</concept_desc>
%<concept_significance>500</concept_significance>
%</concept>
%</ccs2012>
%\end{CCSXML}
%
%\ccsdesc[500]{Security and privacy~Distributed systems security}
%
%\keywords{VoIP security, end-to-end encryption, SAS validation, key exchange validation, mobile app security} % TODO: replace with your keywords

\begin{abstract}

Many widely used Internet messaging and calling apps, such as WhatsApp,
Viber, Telegram, and Signal, have deployed an end-to-end encryption
functionality. 
%The goal is to hide the communications from everyone
%\textit{except} the end users. 
To defeat potential \textit{man-in-the-middle}
attackers against the key exchange protocol, the approach crucially relies
upon users to perform a \textit{code verification} task whereby each user
must compare the code (a fingerprint of the cryptographic keys) 
computed by her app with
the one computed by the other user's app and \textit{reject} the session if the
two codes do not match. 

In this paper, we study the security and usability of this human-centered code
verification task for a prominent setting where the end users are \textit{remotely
located}, and compare it as a baseline to a potentially less frequent scenario where the end users are
in close proximity.  We consider several variations of
the code presentation and code verification methods, incorporated into representative real-world apps, including codes encoded as 
numbers or images, displayed on the screen, and verbally spoken by the users.  We
perform a carefully-designed human factors study in a {lab setting} to 
quantify the security and usability of these different methods.  
%As our
%security metric, we use False Accept Rate (FAR), the rate at which
%non-matching codes are accepted. As our usability metric, we use False
%Reject Rate (FRR), the rate at which matching codes (i.e., valid sessions) are rejected, 
%%and 
%%measure users' delay in performing the tasks, 
%and user
%perception ratings through standard usability questionnaires.

Our study results expose key weaknesses in the security and usability of
the code verification methods employed in the remote end-to-end encryption apps.
First, we show that generally most code verification methods offer \textit{poor
security} (high false accepts) and \textit{low usability} (high false rejects
and low user experience ratings) in the remote setting. 
%Second, we show that methods that have better usability also have poorer
%security, highlighting a classical \textit{trade-off} inherent to a
%human-centered security system. 
Second, we demonstrate that, security and usability under the remote code
verification setting is \textit{significantly lower} than that in the 
proximity code verification setting. We attribute this result to the increased  cognitive overhead  associated with comparing
the codes \textit{across two apps} on the same device (remote setting)
rather than \textit{across two devices} (proximity setting).
%, further emphasizing the intricate challenges associated with securing remote
%communications.
Overall, our work serves to highlight a serious fundamental vulnerability of Internet-based
communication apps in the remote setting stemming from human errors.
%committed by naive users of these
%apps. 

% We conclude our work by
%providing insights for the app designers to lower the impact of the exposed
%vulnerability, and by presenting future directions for the security community.

\end{abstract}
\maketitle

%%TODO 4 apps

\section{Introduction}
\label{sec:intro}

%Mobile instant messaging and Internet calling applications may be quickly replacing
%short messaging (SMS) and calling services provided by mobile network operators.
%%and wireless service providers. 
%The Internet messaging and calling applications
%offer free or low cost services and provide several features traditionally not
%available in SMS messaging and cellular calls. However, a natural concern about Internet 
%communications is the security of the service, since it is commonly believed that Internet is an 
%easier target for the attackers than phone networks.
%%and therefore offers lower security guarantees 
%%compared to the same.

%\smallskip
%\noindent \textbf{End-to-End Secure Internet Communications:} 

Many widely deployed Internet-based messaging and calling applications, such
as WhatsApp \cite{whatsapp}, Viber \cite{viber}, Telegram \cite{telegram} and
Signal \cite{signal}, have deployed an end-to-end encryption (E2EE) feature, to
hide the communications from the attackers and even from the service providers.
Using this approach, all the communication between the end users gets encrypted/authenticated
with a key held only by the communication parties. To share the secret key, the
end users run a key exchange protocol (e.g.,
\cite{zrtp,Vaudenay05,otrv3,secretchat}) over the insecure public Internet
 (or a channel controlled by the application service provider). The key exchange protocol then results in an initial master key that is
used subsequently to generate session keys for encrypting all the messages, including text, data, and voice. 

In contrast to typical client-to-server encryption or PKI-based secure
communication (e.g., TLS), E2EE reduces the unwanted trust
onto third parties (e.g., an online server), because such services may
themselves get compromised, be malicious or under the coercion of law
enforcement authorities.  However, the key exchange protocol runs over
unauthenticated insecure channel and is therefore susceptible to a
Man-in-the-Middle (MITM) attack \cite{mitm-voip, gov3}. To defeat potential
MITM attacks against the key exchange protocol, E2EE apps compute a
readable/exchangeable ``security code'' (a fingerprint of the key exchange protocol) that is used to provide end-to-end
authentication.

The E2EE apps crucially rely upon the users to perform a \textit{code verification} task whereby each user {exchanges and compares} 
the security code computed on her
end with the one computed by the peer's device, 
and must \textit{reject} the session if 
the
codes \textit{do not} match. 
In this context, the failure of the users in matching the codes (i.e.,
accepting {completely or partially} mismatching codes) will be catastrophic in terms of security as it will lead to
the success of the MITM attacker and exposure of all communicated messages.
The users are usually in possession of only their mobile devices
using which the communication takes place, not any other devices or aids (such as paper-pencil), during
this code verification task.

%, and even tampering of text messages by the MITM attacker.  
\begin{comment}
At the same time, the
failure of users in accepting the matching codes (benign, i.e., an MITM attack-free,
session) would lead to the rejection of the communication session and a restart of the
protocol, which will hamper usability
\end{comment}

\smallskip
\noindent \textbf{Remote vs.\ Proximity Code Verification:}
Security code verification has long been used in multiple security
applications, primarily by the users that are located in a \textit{close physical proximity}
%and for the
%security code that are short 
(e.g., in device pairing
\cite{kainda2009usability} and decentralized trust management
\cite{blaze1996decentralized} applications).  However, Internet messaging and voice
applications open up an important area of \textit{remote communications} for people who 
are at distant locations, and 
%may not meet in person prior to communicating with each other using these apps.
have never
met each other or are likely to start communicating using the apps 
before they meet in person. 
 
\begin{figure}[!t]
\vspace{-8mm}
\includegraphics[width=8cm]{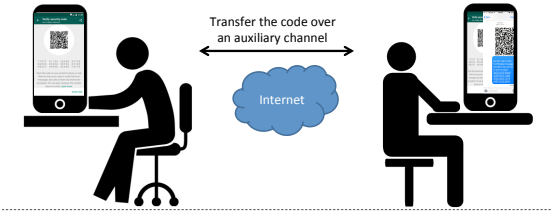}
\vspace{-16mm}
\caption{{Proximity setting (\textbf{cross-device comparison}) }}
\label{fig:local}
\end{figure}

\begin{figure}[!t]
\vspace{-8mm}
\includegraphics[width=8cm]{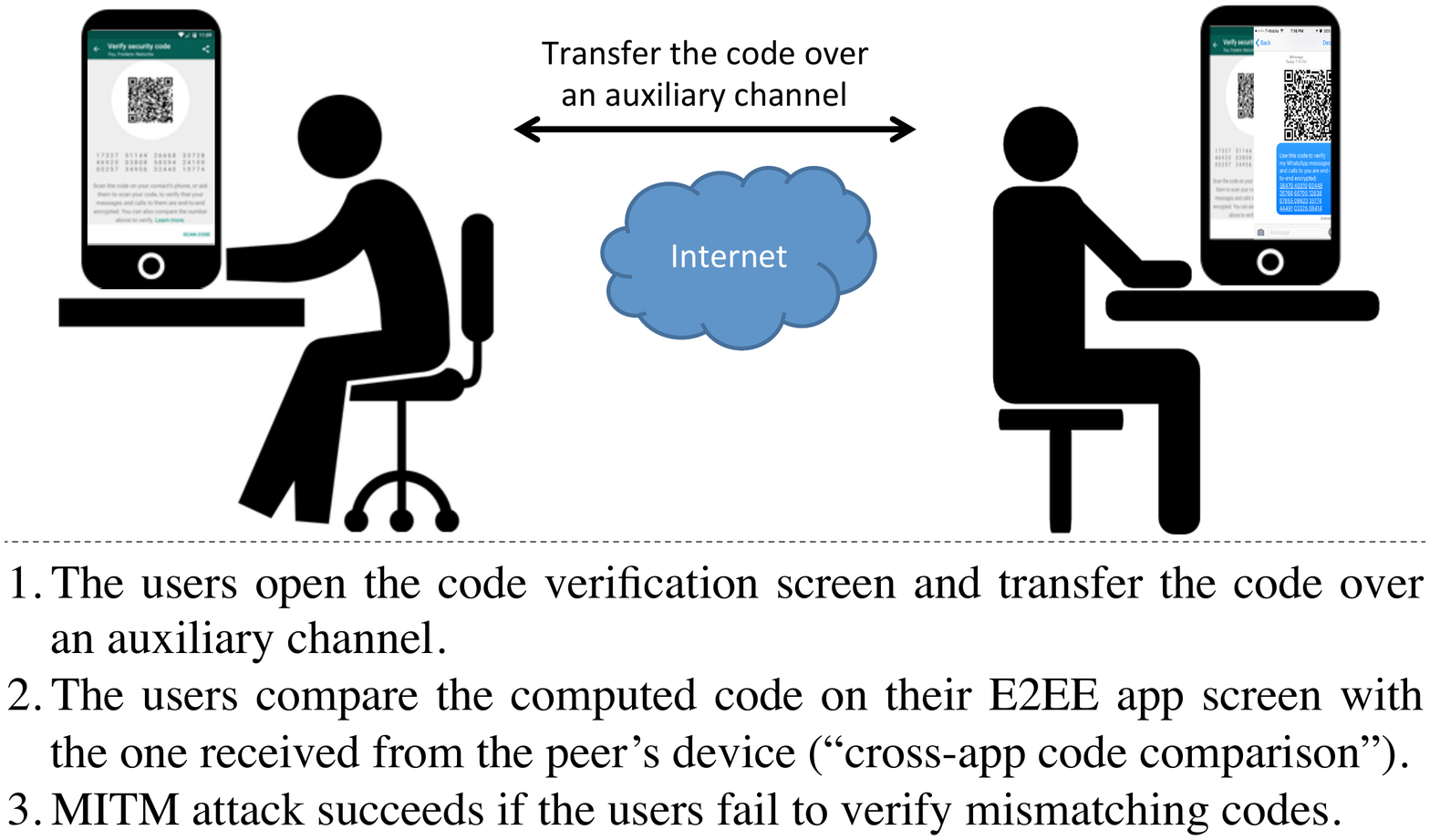}
\vspace{-16mm}
\caption{{Remote setting (\textbf{cross-app comparison})}}
\label{fig:remote}
\end{figure}

%{\color{red}
The remote communication paradigm presents many challenges for code
verification compared to proximity settings,  especially due to the lack of physical
interaction between the users. Therefore, the users are required to exchange
the code over an auxiliary channel (e.g., a voice call or an out-of-band channel such as SMS and email), and then compare the codes \textit{across two
apps} (e.g., an SMS messaging app and the E2EE app) on the same device. Such ``cross-app
comparisons''  might impose cognitive burden on the users, since they may be
required to memorize the received code and compare it with the one displayed on
the E2EE app. In contrast, in the proximity setting, code comparisons seem much simpler 
since the users are close by and compare the codes \textit{across two devices}. Figure \ref{fig:local}
and \ref{fig:remote} depicts the proximity and remote settings, respectively.
%Hence, the current apps need a careful consideration of code
%verification for the remote setting.

\smallskip \noindent \textbf{Our Focus---Security and Usability Study of Remote Code
Verification:} In this paper, we study the security and usability of the
human-centered E2EE code verification specially in a ``remote
setting''. As a baseline, we compare the results with a ``proximity setting''.
%where
%the end users meet in person or are in close proximity. 
Although the
security and usability of the code verification for various security
applications in a proximity setting has been studied in literature before 
\cite{uzun2007usability, kumar2009comparative, kainda2009usability,
hsiao2009study, dechandempirical}, to our knowledge, this
is the first study of the code verification in a potentially more common remote
setting in the context of end-to-end messaging and voice apps. 

We study several remote/proximity code presentation and verification
methods covering approaches deployed by popular E2EE apps, including, WhatsApp
\cite{whatsapp}, Viber \cite{viber}, Telegram \cite{telegram}, Signal
\cite{signal}, Threema \cite{threema}, Wickr Me\cite{wickrme}, and ChatSecure
\cite{chatsecure}.%, and Silent Circle \cite{silent}.  

The founding hypothesis of our study is that remote code verification will
be highly error prone for end users to perform due to the challenges associated
with the remote setting outlines above.
%and the use of relatively long codes in E2EE apps (e.g., 60
%digits). 
Our goal is not to argue on the attackers' ability to manipulate the code
(via either tampering with the key exchange protocol or tampering with the
out-of-band channel) or the amount of the manipulation, but rather to
mainly determine 
%how well the users can verify the matching codes in the benign setting and
how well the users can detect the mismatching codes in the presence of an attack
that can partially manipulate the code. 
 
To test this
hypothesis, we design a human factors study in a controlled lab setting to
measure the accuracy of the users in verifying the security codes.
% in remote setting. 
\begin{comment}
Recall that the decision of the users in accepting or rejecting the codes affects
the security and usability of the system, which is what we aim to investigate 
through our study.
\end{comment}
In the study, we present the participants with several matching codes, representing a
benign, ``attack-free'' scenario, and mismatching codes representing an MITM attack
case.  For the security assessment, we use False Accept Rate (FAR) as our metric specifying
instances of accepting the mismatching codes by the users. Failure of detecting the
mismatching codes indicates the success of the attack. 

For the usability assessment, we quantify user perception through System Usability
%% NS -- add cite to SUS
Scale (SUS) questionnaires \cite{brooke1996sus} and user perception ratings. 
%{\color{red}
As an additional usability
metric, we use
False Reject Rate (FRR) indicating the instances of rejecting the matching codes or the benign case by
the users. Rejecting the benign cases may force the users to restart the
protocol, and therefore, affects the usability of the system as
the process would need to be repeated possibly annoying the users and delaying their
communications.

For our security and usability assessments, we consider several representative remote code
presentations and code verification methods, including, numeric and image codes exchanged over out-of-band
messaging channels (e.g., SMS and email), and verbally spoken codes exchanged
over Internet calls. 
In the proximity setting, we consider the QR, compared almost automatically by the apps, numeric and image code, compared visually by the users.

While our study involves several code presentation and verification methods, the primary goal of our study
is to compare the remote setting with the proximity setting and \textit{not} to compare between different methods.

%In the proximity setting of our study, we consider the QR,
%numeric and image code verification methods, compared almost automatically by the apps (QR code)
%or visually by the users. 

\smallskip\noindent \textbf{An Important Aspect of Our Study Methodology:}
Since our hypothesis is a negative one (i.e., we mainly expect the security of E2EE apps to be relatively poor in the remote setting), we
methodically design our study tailored for the near best defense conditions. 
%We
%essentially provide a counter-example that the E2EE does not work well in terms
%of security and usability. 
To prove our hypothesis, we recruit 25 young,
educated, and technology-aware participants with reasonably good computer and
security background to perform the code verification task willingly and diligently, and with
full awareness of the security task, in a
controlled observable lab environment. Given that in the existing
applications, the code verification task is optional, in practice, the attack
might be more difficult to detect, especially
for \textit{average users} who are not informed about the security risks of
neglectful code verification, and may often skip or click through the code
verification procedure.  
Moreover, in our study, we set the code verification task as the participants' 
only primary task. In real-world, in contrast, the code verification 
task will be a secondary task, the primary task being setting up the communications.
Thus, if our study participants do not perform well with a single task at hand, the 
real-life would potentially do even worse since they have to manage two tasks whereby
the security task may not be the high-priority task.

This study design is in contrast to the
traditional usability study designs where the goal is to demonstrate that the
system is secure or usable, in which case there is a need to prove
security/usability by recruiting a large sample of users with average
technical skills and emulating real-world conditions as much as possible.
%This design can itself be used independently to assess the security of systems. 
%
%That is, we recruit young, educated, and technology-aware participants with good computer and security background, who perform the task willingly and diligently.  Given that in the existing applications, the code verification task is optional, in practice, the attack might be more difficult to detect for average users who are not informed about the security risks of neglectful code verification, and often skip or click through the security procedures.

%%Results

\smallskip \noindent \textbf{Our Primary Results:}
Our results are aligned with our hypothesis, and show that
the code verification methods deployed in E2EE apps suffer
from several security and usability issues arising from human errors in
verifying the codes in remote settings: 
%The key results are:
%summarized below:

\begin{enumerate}[leftmargin=*]

%%Quantify
	\item \textbf{\textit{Low Security for Remote Code Verification}:}
		All the remote code verification methods have high FARs,
		ranging on average from about \textit{13\% for image code verification}
			to \textit{40\% for numeric code verification.} 
		{Further, if the attacker has more control over the number of matching characters between an attacked code and a legitimate code, the success rate of the attack would increase.
		For instance, the FAR \textit{increases to about 60\%} in numeric code verification, when only one digit is mismatching between the codes. }
		These error rates are
		exhibited by young educated participants in a unidirectional
		authentication setting. In practice, users with more diverse
		background may perform worse than the participants in our
		study. 
		Also, the error rate increases to \textit{almost double} in a
		\textit{bidirectional authentication} setting, where the attacker may
		deceive any of the two parties in an E2EE secure channel
		establishment session.  
		\vspace{1mm}

	\item \textbf{\textit{Low Usability for Remote Code Verification:}} The results also point out the usability issues with the remote code verification 
methods. Except for the audio-based code verification, \textit{the other remote schemes have about 20\% FRR}.
%% NS 2/11: can we quantify the SUS and other ratings?
Further, in terms of system usability scale and the users' personal ratings, results are indicative of a poor user experience underlying the remote setting 
(e.g., \textit{SUS scores around only 50\%}).

\begin{comment}
%\vspace{2mm}
\item \textbf{\textit{Security and Usability Trade-Off in Remote Setting:}} We observed that the methods, such as image code verification, that have lower FAR (about 13\%), exhibit a high FRR or low usability (SUS score of about 45), whereas schemes, such 
as numerical code verification, that have high FAR (about 40\%), yield a lower FRR or a relatively higher 
usability score (SUS score of about 54). This result highlights the classical security-usability trade-off inherent to a
human-centered security system in general and E2EE apps in particular. 
\end{comment}

		\vspace{1mm}
\item \textbf{\textit{Remote Setting vs.\ Proximity Setting:}} As our baseline condition, we measured the
	security of the code verification in a proximity setting, which shows that
	users could successfully detect the benign and attack settings with
	negligible error rates.  However, in a remote setting which is the
	primary model of the E2EE apps, {the error rates---FAR and FRR---are
	(statistically) significantly higher}. Moreover, the user perception ratings in the
	remote scenario were (statistically) significantly lower compared to the proximity
	setting.

\end{enumerate}

%insights

\noindent \textbf{Generalizability and Broader Impacts:}
Overall, we believe that our work  highlights a serious fundamental
vulnerability of a broadly deployed and extensively used representative class
of secure Internet-based remote communication apps.  This vulnerability does
not arise from the underlying cryptographic protocol or the software
implementation of the protocols, but rather from the human errors committed by
naive users of these apps. It mainly stems from the fact that, in the remote
setting, comparisons of security codes need to be performed across the apps,
which is burdensome for the users and thus highly error-prone.  Such cross-app
comparisons also significantly impact the usability of the code verification
process.  Although our study focuses on many currently deployed apps for a
larger impact to real-world systems, our work is generalizable in that it shows
the security-usability pitfalls of cross-app comparisons broadly applicable to
the fundamental design of remote E2EE code verifications.

Addressing this vulnerability and
related usability issues will be a challenging problem, but we hope that our
work will encourage the app designers to make necessary changes to their systems to
improve the levels of security and usability in the face of human errors. 
%{\color{red}Although we do not specifically target other security applications of hash/fingerprint verification (e.g., GnuPG \cite{callas2007openpgp}), our study could be extended to such applications.}
%\begin{comment}
Based on the results of the study, we also provide insights and future
directions for the designers of the apps to improve the security and usability of their systems, without
affecting the current threat models.  
%We also suggest using multi-screening options offered by some smartphones  
%\end{comment}

\begin{comment}
%Our preliminary study of this 
%defense mechanism is positive and promises higher security and usability.
\smallskip
\noindent \textbf{Paper Outline:}
In Section \ref{sec:back}, we discuss the E2EE protocol and the code verification methods and briefly review related works. In Section \ref{sec:preliminaries}, we define objectives, hypotheses and assumptions of our study. The design of the study is introduced in Section \ref{sec:design}, and the result of the user study is discussed in Section \ref{sec:eval}. In Section \ref{sec:discuss}, we  discuss the insights from our study and future directions. 
\end{comment}

%% TODO

\section{Background}
\label{sec:back}

%\subsection{End-to-End Encrypted IM and VoIP Apps}
\subsection{End-to-End Encryption Apps}

%To prevent eavesdropping of the communications, E2EE is used to ensure that only parties involved in the communication can encrypt and decrypt 
%the conversation. 
Recently, several instant messaging (IM) and Voice over IP (VoIP) applications  adopted 
the E2EE protocol %in the context of secure instant messaging and VoIP 
to provide 
secure communication of messages/calls.
%over the insecure communication 
%channel. 
Examples of such applications are Signal \cite{signal}, WhatsApp 
\cite{whatsapp}, Facebook Messenger \cite{messenger}, Google Allo \cite{allo}, 
Viber \cite{viber}, Telegram \cite{telegram}, and Threema \cite{threema} (a more 
comprehensive list can be found in Appendix \ref{app:apps}). 
%In these 
%apps, all messages, including call, text, photos, and files, can 
%be encrypted with the keys held by only the conversation 
%participants. 
%*****For the sake of simplicity we assume only end-to-end encrypted text 
%messaging in this section. 

Different E2EE protocols have been proposed and implemented by 
the apps targeting the specific needs and challenges facing the IM and VoIP
communications. Off-the-Record Messaging (OTR) 
protocol \cite{borisov2004off} by Borisov et al. is an E2EE
protocol specifically designed for IM communication, with the goal of perfect 
forward secrecy (PFS) through ephemeral key exchanges to avoid long-lived keys, 
authenticity of the parties, non-deniability of the messages, and 
confidentiality through encryption.

Many of the current E2EE applications are constructed on top of 
OTR, however, they adapted OTR to work with asynchronous transport (which is  
specific to the smartphone environment) \cite{otrv3}. 
The encryption protocol introduced by Open 
Whisper System is one of the leading protocols
 that was first developed by Signal \cite{signal}, and later 
implemented in WhatsApp \cite{whatsapp}, Facebook Messenger \cite{messenger}, 
and Google Allo \cite{allo}. 
%Signal protocol (a.k.a 
%Axolotl Protocol, and TextSecure Protocol)
The protocol is based on OTR  combined with Silent Circle Instant 
Messaging Protocol (SCIMP) \cite{SCIMP} for improved forward secrecy. The 
protocol is also enhanced to allow asynchronous messaging, and to simplify 
non-deniability
%. The detail of Signal encryption protocol can be found in 
\cite{doubleratchet}.  
Viber \cite{viber,viberenc} %follows similar approach and 
%as the one introduced by Open Whisper System, however, the 
%implementation does not borrow the code from Signal. 
and Telegram \cite{telegram,secretchat} have also adopted similar protocols. 
%called MTProto Mobile Protocol 
%for its E2EE ``Secret Chat'' 
%(for details please 
%refer to 
%\cite{secretchat}. 

\subsection{MITM Attack Resistance}

The encryption key in the E2EE applications is typically exchanged between the
endpoints during an initial key exchange protocol over the insecure channel (or
a channel fully controllable by the service provider) and is then refreshed
during the message exchanges.  The basic key exchange protocol is based on the
well-known Diffie-Hellman (DH) key exchange protocol \cite{dh}. 
%% next paragraph can be removed
%DH assumes the two large prime number $p$, and $g$ as the public parameters.
%The two parties in the protocol (Alice and Bob) each picks a random private
%key (Alice picks $a$ and Bob picks $b$) and sends the corresponding public key
%$g^a$, and $g^b$ to the other party, the shared key can then be computed as
%$g^{ab}$.  E2EE reduces the trust onto third parties (e.g., certificate
%authorities), which might themselves be compromised. 
However, the basic Diffie-Hellman key exchange protocol is not authenticated
and therefore is susceptible to the MITM attack. Hence, the applications add a
manual  authentication scheme, which we refer to as ``code verification''  to
the key exchange to protect against MITM attack. 

Regardless of the differences in the E2EE protocols developed by different
apps, they all rely on the human users to perform this code verification task in
an attempt to defeat the MITM attack. This human-centered task is the focus of
our work.

\subsection{Code Generation and Presentation}

For authentication, users can compare the public keys (or a fingerprint/hash of the 
public keys) of other users over {an auxiliary channel  (e.g., an out-of-band  channel such as text or email, or a human voice channel). }
%to verify each other's identity
To simplify the manual comparison, the public key or the fingerprint is encoded into a 
readable/exchangeable code. 
%The users of the apps are expected to compare 
%the code {to verify each other's (or their device's) identities to make sure that the key exchange protocol is not susceptible to MITM attack.} 
%Different apps have different terminology to refer to these codes (e.g. safety 
%numbers, identity, security code, fingerprint). 
We 
refer to this code as ``security code'' or in short ``the code''. 

%\vspace{-2mm}
%\subsubsection{Code Generation}
% 
%
% 
%%Here we summarize some of the common code generation, encoding and verification methods.
%
%Usually the code is derived from the public key of the users by 
%hashing the public key, and truncating it into a $t$-bit code. For example, the 160-bit output of SHA-1 of the public key can be truncated to a 30-bit code. 
%%Details of the code generation of each and every app is not in the scope of this 
%%paper and can be found for different apps by referring to the technical 
%%documentation of the encryption protocols. However, 
%In section \ref{sec:app}, we 
%briefly introduce the code generation for the E2EE apps we used in our user study.%, namely, Signal, Telegram, Viber, and WhatsApp. 

%\subsubsection{Code Presentation}

Specifically, the generated code is presented as a human readable code or an exchangeable  
object.  Often a long code is represented as blocks (or chunks) of few digit numbers 
for readability (e.g., 30-digit  presented as 6 chunks of 5-digit numbers).
Figure \ref{fig:presentation} in Appendix \ref{app:tabs} shows some of the common code presentations described here: 

\begin{itemize}[leftmargin=*]
\item\textbf{{QR Code}}: The code is encoded into a QR code. Since  the QR code 
is automatically captured and compared by the app without relying on human user, 
it can contain the full public key of the 
parties. Figure \ref{fig:whatsapp} shows the QR code presentation for WhatsApp, 
which includes a version number, the users' identifier, and the  
public key for both parties. 
\item\textbf{{Numeric}}: The code is presented as a sequence of numeric digits.
 WhatsApp, Signal, Telegram, and Viber
use this form of presentation.  Figures \ref{fig:whatsapp}, \ref{fig:signal}, and \ref{fig:viber} show the code in 
numeric presentation in WhatsApp, Signal, and Viber application. 
%\item\textbf{{Hexadecimal}}: The code is presented as a sequence of Hexadecimal numbers 
%(0-9, a-f). Threema, Telegram, and earlier version of Signal use this type of presentation. 
%Figure \ref{fig:signal} shows the code in hexadecimal presentation in earlier 
%versions of Signal.
\item\textbf{{Images}}: The code is encoded into an image. Examples of such images are 
also presented in prior literature (e.g., \cite{perrig1999hash, dhamija2000hash}). 
Telegram encodes the code in a shaded blue square as 
shown in Figure \ref{fig:telegram}.
%\vspace{2mm}
%\item\textbf{{Words}}: The code is presented as typically a short string (2-4 words). PGP 
%words \cite{juola1996whole} is an example of this presentation in which one byte of the code is encoded into a word. Silent Circle uses a 2-word PGP word to represent the code as shown in Figure \ref{fig:silentcircle}. Since our study only involves long codes, we do not consider this presentation in our study. 
\end{itemize}

Usually a set of one or more of the mentioned presentation is used by the apps. 
Other encodings such as {words \cite{juola1996whole}}, alphanumeric, Base32 and Base64, sentences, and phrases, 
have also been proposed in literature, however, they are not being used at this moment by any of {the popular E2EE apps we study in this paper}. 

%%%PROXIMITY Setting rather than remote setting
%\subsubsection{Proximity vs. Remote Setting}

%The users who are involved in the code verification are either in close proximity 
%(e.g., meeting in person) or in two different locations. We refer to the former
%setting as ``proximity setting'' (depicted in Figure \ref{fig:local}) and to the latter as ``remote setting'' (depicted in Figure \ref{fig:remote}). 

%%EXPAND

%Proximity setting is typically used for device pairing, while remote setting is
%more common for IM and VoIP applications that are designed for remote
%communications.  Since in the proximity setting the users are nearby and can
%see each other, the code exchange channel is by default authenticated and simple to perform.
%However, in the remote setting, the codes should be exchanged over an {auxiliary
%	channel and
%	then compared \textit{across two apps} on the same device.
%	Such ``cross-app comparisons''  might impose cognitive burden on the
%	users, since they are required to memorize the received code and
%	compare it with the one displayed on the E2EE app. 

%	In contrast, in the
%	proximity setting, code comparisons seem much simpler since the users
%	are close by and have to compare the codes \textit{across two device}.

\subsection{Code Exchange and Verification}

There are two notable approaches to verify the security codes in security applications, Copy-Confirm and 
Compare-Confirm as introduced in \cite{uzun2007usability}. 
In Compare-Confirm, the code is displayed on each peer's
screen, they exchange their respective code, 
and both accept or reject the connection by comparing the displayed and received 
code. In
Copy-Confirm, one party reads the code to the other party,
who types it onto his/her device, and gets notified whether the code is
correct or not.
The E2EE apps follow the Compare-Confirm.
%Compare-confirm is the prominent verification method for words 
%and 
%long codes 
%as is the case for the E2EE apps. 

In a proximity setting, where the two parties are physically co-located, 
the code verification can happen by visually matching the  
codes (e.g., comparing the 60-digit numerical code in WhatsApp, or comparing the 
graphic in Telegram app). 
%%MSH should justify the relation between numeric and QR
Some apps provide automated code comparison by embedding 
 the code in a QR code that can be captured and compared by the app. 
%(this approach can be considered as an automated Copy-Confirm code verification approach).

In a remote setting, the two parties can exchange the 
codes over an out of band channel, for example, through a text message, or an 
email. This approach is very common among several apps such as Whatsapp,  and Signal.
%, and Threema. 
Another method of code exchange in a remote setting 
is transferring the code over an 
authenticated channel, for example, through a voice call (e.g., used by Viber). This approach assumes that the human voice 
channel over which the code is exchanged is authenticated. 

\subsection{Threat Model}

In this work, we study the common threat model followed by the E2EE
applications. It is a setting where two trusted parties, Alice and Bob, intent
to start a secure communication over an insecure channel using only their mobile devices.
In this threat model,
Alice's and Bob's devices are trusted. However, the communication channel over
which the key is exchanged is assumed to be fully controlled by the attacker.
That is, the attacker can intercept and modify the key exchange messages. 

The key exchange protocol is followed by an authentication phase to defeat MITM attacks. 
%MITM attack refers to the case where the attacker can tamper with the protocol messages. 
In the proximity setting, the code exchange channel is physically-authenticated since the users
are in close proximity and can verify each other's identity. In the remote setting, the channel over which the code is transferred should somehow be authenticated. 
An example of a supposedly authenticated channel is out of band messaging (e.g., SMS, email or social media). 
Another authenticated channel used by the applications is voice channel. The assumption is that the users can recognize each other's voice, and therefore, can authenticate each other when the code is spoken. However, as we discuss below, even the out-of-band or voice channels may be tampered with, at least partially. 

%
%We consider the remote setting as the primary model, while we define the local
%setting as a baseline for the security and usability of the code verification. 
\smallskip
{ \noindent \textbf{Attacker Capability:} We consider two levels of capability
	for the attacker. First, we assume that the attacker does not have any
	control over the out-of-band or voice channel.  This assumption is
	generally made by the app designers and motivates them to transfer the
	code over such channels.  Even though the attacker has zero control
	over the channel, she can interfere with the key exchange protocol and
	run a preimage attack\footnote{A preimage attack on hash functions
	tries to find a message with a specific hash.} on the
	security codes to generate attacked codes that are the same or at least
	partially similar to the legitimate security codes (in a hope that the user
	accepts it due to similarity with the legitimate code).  The attacker's goal
	is to guess the public key pairs that generate codes with maximum
	similarity to the protocol's legitimate security code. However, the computational
	cost limits the success level of the attack.  This level of capability
	is considered for the attacker in several prior studies about public
	key fingerprint security \cite{dechandempirical, unicorns}. 

Second, we assume that the attacker has control over some part of the messages
transferred over the out-of-band or voice channel.  That is, the attacker may
be able to tamper with the exchanged codes partially.  The
attacker first interferes with the key exchange protocol, which results in two
different codes at the two parties. Then during the code exchange, the attacker
tampers with the messages transmitted over the  out-of-band or voice channel by
inserting or modifying one or more bits of the transmitted code. For example,
if $code_A = 1234$ and $code_B = 4567$, the attacker may change $code_B$, while
it is being transmitted to Alice, to $1534$ (a code that is different from
$code_A$ in only 1 digit).

As an example for the voice-based channel, the attacker may have access to some pre-recorded samples of the
user's voice, speaking some of the digits that appear in the code (1, 3,
and 4 in $code_A$, but not 2). The attacker then tampers with the voice messages of 
Bob, while Bob reads the code to Alice, and
injects those pre-collected samples (i.e., change 4 to 1, 6 to 3, and 7 to 4) so that 
$code_B$ received by Alice is $1534$. Such an attack is known as a voice MITM
attack \cite{ccs-2014}, which allows the attacker to tamper with the voice channel by
mimicking users' voice or playing the pre-recorded samples. 

Another instance of such part manipulation pertains to altering the codes transmitted over the text-based channels,
such as SMS, emails, or even social media posts. Most of the E2EE apps allow
exchanging and sharing of the codes on different messaging and social media
applications. This gives the attacker the opportunity to compromise any of these
applications or the user accounts to modify the exchanged code. 
Also, email and SMS spoofing\footnote{e.g., a website, \url{http://www.smsgang.com/}, can be 
used to send spoofed SMS messages.} is a common problem that would make
such attacks highly feasible in practice.
Since some of these apps have a character
limit (e.g., Twitter and SMS apps), they may not be able to accommodate the code in one message, and
therefore have to split the code into few parts. If the attacker is able to modify
some of these fragmented parts, the attacked code appear to be similar to
the legitimate code. 

%% NS 2/12: the first sentence can be futher clatified -- I tried but you should check
In summary, we allow these two levels of capability for the attacker, that is,
the attacker may have zero or partial control over the code, and can modify it
partly via a pre-image attack indirectly or via tampering with the auxiliary
channel directly.  It is worth noting that the attacker who has \textit{full
control} over the code may change the code such that it completely matches a
legitimate code and be impossible for the users to detect.  In this light, such
a full manipulation attack is not the focus of our study, since this attack may
have 100\% success rate (unless a user mistakenly rejects this ``attacked''
matching code, which as our study shows happens up to only about 20\% of the times).

%%TODO
%% add apps version number 
%% add number of bits in the code

\section{Study Preliminaries and Goals}
\label{sec:preliminaries}

\subsection{Study Objectives}
\label{sec:objective}

%Our study is designed to evaluate the security and usability of E2EE remote code verification. 
The specific goals of the study are outlined below:

 \noindent  \textbf{Robustness:} \textit{How accurate are the users in verifying the code?}
Robustness directly impacts two important properties of the system: security and usability. 
 
 \begin{itemize}[leftmargin=*]
  \itemsep0em
 	
	\item \textit{For security assessment}, we are interested in
	determining how often users accept mismatching security codes.
	\textit{False Accept Rate (FAR)} denotes the probability of incorrectly accepting
	such instances. False acceptance implies the success of the MITM
	attack and the compromise of the  system. 
	
	   \item \textit{For usability
	assessment}, we are interested in finding out how often users reject
	matching security codes. \textit{False
	Reject Rate (FRR)} represents the probability of incorrectly rejecting benign instances. False rejection forces the users to restart the
	protocol affecting the overall usability. 
 \end{itemize}

%\item \textbf{Efficiency:} \textit {How long does it take the users to verify a code?} The delay may impact the overall usability of the system since the primary task of establishing a secure conversation may prolong.
  \noindent \textbf{User Experience and Perception:} \textit {How usable do the users find the code verification task.} 
We define the following parameters to measure the usability of the system:  

\begin{itemize}[leftmargin=*]
  \itemsep0em
  \item{\textit{System Usability Scale (SUS):}} How easy or difficult do the users find the system? Can the user easily learn how to use the system? Do they need the support of a technical person?
  \item{\textit{Comfort:}}  How comfortable are users with the system? 
  \item{\textit{Satisfaction:}}  How satisfied are the users with the system in verifying the codes?
  \item{\textit{Adoptability:}} Are they willing to use the system in real-life?

\end{itemize}
%  \textit{Trust:} Do the users trust the system?

%%/*****remove trust from the objectives  

%\end{itemize}

\subsection{Selected Applications}
\label{sec:app}

%Our main criteria in selecting the applications is the E2EE feature whether it is enabled on all communications (e.g.,  Signal) or offered as an optional feature (e.g., Secret Chat of Telegram). 
%We picked apps that at least provide text messaging while some of the apps might offer several other features such as voice call, photo sharing, and file transfer. 
%Since we would like to study the code verification method, 
We selected a collection of highly popular and representative apps, based on the \textit{number of installations and ratings}, to comprehensively cover different code verification methods including QR, textual, image, and voice verification. 
In the textual presentation, we only picked numeric presentation, given that it is the most commonly adopted method.
%compared to words, phrases, and sentences. 

To cover all the state-of-the-art code presentation and verification methods, and based on the popularity of the apps, we picked WhatsApp {(v.2.16.201}), Viber {(v.6.2.1251}), Telegram {(v.3.10.1}), and Signal {(v.3.15.2}) in our study. The complete list of studied apps is described in Appendix \ref{app:apps}. 

%Also we believe
%%%MSH is the next sentence correct?
% this is a fair decision since words and sentences are usually short codes while the numeric presentation is longer. Short codes provide a lower theoretical security compared to longer codes, and are perhaps more usable 
% that the longer codes. 
%%%MSH Should we include this comparison? 
% However, in Section \ref{sec:discuss} we provide a comparison with the reported results from \cite{acsac-2015} who evaluated the security of the Short Authenticated String (SAS) comparison in the presence of the data and voice MITM attacks on E2EE VoIP applications.  

%%Change the ordering based on the app popularity
\noindent \textbf{App \#1---WhatsApp:}
WhatsApp displays the code as a QR code and a 60-digit number represented as 12 blocks of 5-digit number. The QR code includes a version number, the user's identifier, and the identity key for both parties.
WhatsApp generates {its 60-digit (480-bit code} by concatenating the 30-digit fingerprint of the users' 
public Identity Keys\footnote{Identity Key is the long-term Curve25519 key pair 
generated at install time}. To calculate the fingerprint, iteratively SHA-512 of 
the public key is computed for 5200 times and the first 30 bytes of the final 
result is output as the security code.  
%Figure \ref{fig:whatsapp} shows the 
%presentation of the code. 
%This approach is used in recent releases of Signal as well. 
  
Users have two options to verify each other's identity. First, they  can scan the QR code of the peer user; the app then compares and 
verifies the codes. This feature is helpful if the users are in close proximity. 
Second, the app allows the users to exchange the code through email or messaging applications and then manually verify the code. This feature can be used in the  remote setting.

 \noindent \textbf{App \#2---Viber:}
In Viber, both devices perform DH key exchange using their own private keys and the peer's public key.  
The DH result is hashed using SHA256 and {trimmed to 160 bits}. The code is encoded as 48 digits, displayed as 12 blocks of 4-digit numbers. 
%Figure \ref{fig:viber} shows the presentation of the code. 

To verify the code, the two parties should make a Viber call, during which they
can open the verification screen. The users verbally announce the displayed
code to each other and validate the identity of the peer if the spoken code and
the displayed code match. Viber does not support any other out-of-band or
automated code verification method. 

\noindent \textbf{App \#3---Telegram:}
Telegram generates the code by using the first 128 bits of SHA1 of the initial key (generated when Telegram Secret Chat is first established) followed by the first 160 bits of SHA256 of the key used when the Secret Chat is updated to layer 46\footnote{In Telegram, the layer number is initialized at 8 when the Secret Chat is created and is updated after receiving packets (for details, see https://core.telegram.org/api/end-to-end.)}.
The code is displayed to the users as a shaded blue image along with a string of 64 hexadecimal characters. 
%The users are supposed to 
%compare the code in a proximity setting or through an out of band channel.  
Telegram does not provide a facility to send the code directly from the app.  
%(as shown in Figure \ref{fig:telegram})

 \noindent \textbf{App \#4---Signal:} 
The version of Signal used in the study displays the full 256 bits of the public key hash as a hexadecimal string and a QR code. The users can scan the QR codes to verify each other's identity in a proximity setting or transfer it over an out-of-band channel in a remote setting. 
%Recent version of Signal acts similar to WhatsApp, however, similar to Telegram it does not provide a facility to transfer the code directly from the application. 

%(as shown in Figure \ref{fig:signal})
\vspace{-2mm}
\subsection{Study Assumptions and Hypotheses}
\label{sec:hypothesis}

Our hypothesis is that the apps offer a low security level, due to human errors, in verifying the code and that the users find the code verification to have poor usability.  

Since our hypothesis is negative (apps provide a low level of security and poor usability), we design our study tailored for the near best possible success of the attack. 
We assume
that the applications inform the users about the E2EE feature
and importance of the code verification task. We also assume
that the users are required to perform the code verification
task. Also, we target young, and educated users who are
technology-aware. We also assume that the users perform
the code verification task whether they are planning to start
a conversation with a peer or not. This assumption implies
that users consider the code verification task as a primary
task. In practice, the primary task of the users is to pursue
a conversation and only the secondary task is to verify the
code.

All these assumptions make the attack more difficult since: 
(1) the users are well-informed about the security risks associated
with incorrect code verification, (2) the users are enforced and
willing to verify the code, (3) the users perform the task carefully,
and (4) they are only focused on a single task of code
verification.

%
%Considering the above assumptions we expect the MITM
%attack to be less probable to succeed in real-life, since 
%are expected to perform better than average users of the apps
%who may not be well-informed, young, and educated, and may
%skip the code verification or perform it neglectfully.

%We conduct the study in a controlled lab environment where the participants are 
%expected to perform better than real life due to Hawthrone effect. 

%%TODO
%% empty citation
%%%%This approach is inline with other device pairing studies \cite{}.   %%%CITE PRIOR DEVICE PAIRING
%hawthrone effect

\vspace{-2mm}
\section{Study Design}
\label{sec:design}

Our study was designed to evaluate the security and usability of the code verification according to the objectives defined in Section \ref{sec:objective}.
% in the context of popular and representative applications installed on smartphones in an observable and controlled environment. 
The study was approved by our Institutional Review Board and standard ethical procedures were followed, e.g., participants being informed, given choice to discontinue,
and not deceived. 
%In this section, we introduce the study design and setup.  

%Our system emulates Crypto Phones unidirectional
%call establishment/authentication (such as a scenario where a
%customer service call is being made by a user). Our results, however,
%easily extrapolate to a bidirectional setting, as discussed later
%in Section 5.1.

%\subsection{Lab Study}

%%code verification=authentication

\vspace{-2mm}
\subsection{Code Verification Challenges}

We considered two types of code verification challenges in the study, representing benign and attack cases:

%\begin{itemize}[leftmargin=*]
%\itemsep0em
%\item

\noindent\textbf{{Mismatching Codes:}} For each set of code verification tasks, we defined three instances of mismatching code pairs representing the attack setting\footnote{Implementation of the MITM attack is not within the scope of this work and all of the codes were generated manually to emulate the attack.}. %We are interested to know whether the participants can identify the attack successfully. 
%Instances of accepting a mismatching code is measured by the FAR as mentioned in Section \ref{sec:objective}. 
To study the accuracy of the participants in recognizing the attacked session, except for Telegram app\footnote{Since we did not have access to the Telegram image generation tool and the encoding from the fingerprint to the image was unknown to us, we do not have control over the number of incorrect/mismatching characters.}, we created a code with only one mismatching character, a code with one block of mismatching characters (the position of the mismatching character/block(s) was randomly selected), and a completely incorrect code.  {We recall that the threat model allows partial manipulation of the codes through the preimage attack or access to part of the messages transferred over the auxiliary channel. } 
%%%%This approach is inline with other device pairing studies \cite{}.   %%%CITE PRIOR DEVICE PAIRING
%Our hypothesis is that as the number of incorrect characters increases the participants can more easily detect the attack. 
%%%CITE PRIOR DEVICE PAIRING
%\item

\noindent \textbf{{Matching Codes:}} For each set of code verification tasks, we had two instances of matching codes representing the benign case. 
%We are interested to know whether the participants can correctly verify a benign session. Instances of rejecting a code in a benign setting is measured by the FRR as mentioned in Section \ref{sec:objective}.

%\end{itemize}

\subsection{Experimental Setup and Tested Schemes}

To support our study, we designed a survey using the LimeService platform. The survey presented all the information and instructions to the participants and guided them to complete the tasks.  
%All answers  
%%the questions and challenges 
%were stored on the LimeService for further analysis. 

To perform the tasks, each participant was provided with an Android phone. The four applications (WhatsApp, Viber, Telegram, and Signal) were setup on the phone and five contact names and numbers, associated with the study examiner, were stored in the contact list. 
Android is one of the most popular platforms and all of our participants perfectly knew how to operate the smartphone.

The examiner (we call him Bob) acted as the peer in the communication set-up only to read, transfer, or to show the security codes to the participant (we call her Alice). 
%The examiner holds multiple phones to act as different contacts whom the participant should verify. 
In the benign case, Bob displayed or read the correct codes, while in the attack case, he showed incorrect codes. 
%The examiner also observed the study and took notes of the participants performance and questions. 

In the proximity setting, Alice and Bob sat next to each other.  The proximity setting consisted of three types of code verification methods summarized in Table \ref{tab:verify} (Appendix \ref{app:tabs}):
% (labeled as L1-QR, L2-Image, and L3-Numeric) and described here: 
 
\begin{itemize}[leftmargin=*]
\item\textbf{{P1-QR -- Signal QR Code Verification}}:  In this task, Alice opens the code verification screen for each contact on the Signal app to capture the QR code for Bob. The app then compares the internal copy of  Alice's code with the  captured one. 
%by the QR code reader. 
\item\textbf{{P2-Image -- Telegram Image Code Verification}}: Alice and Bob open the code verification screen for a particular conversation. Bob shows the image to Alice who should compare and verify the image.
\item\textbf{{P3-Numeric -- WhatsApp Numeric Code Verification}}:  Alice and Bob open the code verification screens for a particular conversation. Bob shows the screen to Alice who observes and verifies the code. 
\end{itemize}

In the remote setting, Alice and Bob sat in two different rooms to simulate a remote communication scenario. The remote setting consisted of three types of code verification methods (summarized in Table \ref{tab:verify}, Appendix \ref{app:tabs}):
% (labeled as R1-Audio, R2-Image, and R3-Numeric) and described here: 

\begin{itemize}[leftmargin=*]
\item\textbf{{R1-Audio -- Viber Audio Code Verification}}: In this method, Alice calls the given contacts on the Viber app. Bob picks up the Viber call, and the two open the code verification screens. Then, Bob speaks the code and Alice compares the spoken code and the displayed one. 
\item\textbf{{R2-Image -- Telegram Image Code Verification}}: 
	In this method, Bob sends the image in a text (SMS) message to Alice. Alice compares the code on the Telegram screen with the code on the text message screen by switching between the two apps. That is, Alice needs to open one app to memorize the image code, then move on to the second app and compare the code. 
%Perhaps Alice needs to repeat this process multiple times and each time memorize a section of the image. 

%We refer to this method as parallel windowing.
%
%Some of the smartphones allow the user to split the phone screen into two section and open one app in each portion of the screen. Using this feature for code verification, the user can open code verification screen in one portion of the screen and the text message or email app in another portion of the screen. This option would perhaps be easier for code verification, since the user can see both codes at once. Since this option is not available on all smartphones, we only examined the code verification with the parallel windowing in our user study.

\item\textbf{{R3-Numeric -- WhatsApp Numeric Code Verification}}: 
This type of code verification works similar to the image code verification method. The only difference is that the numeric code is exchanged in a text message or email, and Alice  compares the numeric code on the WhatsApp screen with the code on the text message screen by switching between the two apps.

\end{itemize}

\begin{figure}[t]
  \vspace{-6mm}
    \hspace*{6mm}\includegraphics[width=0.5\textwidth]{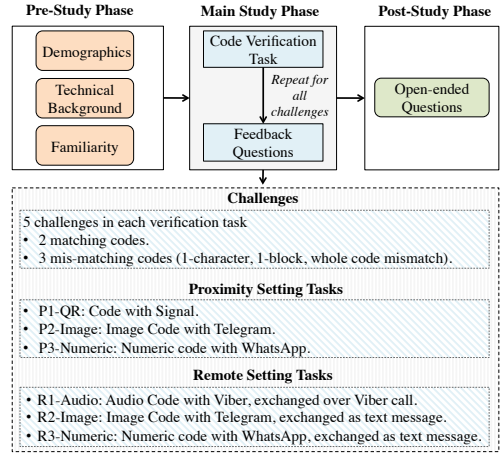}
    \vspace{-10mm}
  \caption{{Protocol Flow of the user study.}}
      \vspace{-6mm}
\label{fig:study}
\end{figure}

\subsection{Study Protocol}

Figure \ref{fig:study} provides the high-level design protocol flow of the study. After a brief introduction to the study, the participants were navigated by the examiner to a computer desk that had the survey web-site open, and were provided with a smartphone.
% We informed the participant about E2EE feature and the claims that the apps make about the security provided by the E2EE. We also warned them about security risks of accepting incorrect codes. 
The participants were informed about the E2EE feature and the risk of accepting incorrect codes. 
%%%%%***** add the info here
They were asked to follow the instructions on the survey website and perform the tasks diligently as presented on the website. 
%%MSH should remove the open-ended questions
The study was composed of three phases: the pre-study, the main
study, and the post-study phase.

\subsubsection{Pre-Study Phase}
The pre-study phase consisted of several qualitative and quantitative questions grouped into three categories, as summarized below: 
%(the complete list of questions can be found in  Appendix \ref{app:pre-study})

\noindent\textit{\textbf{Demographics}:} The participants were asked to fill out a demographic 
questionnaire. These questions polled for each participant's
age, gender and education.  

\noindent\textit{\textbf{Technical Background}:} The participants were asked about their general computer and security skills to uncover their initial attitude towards security. 

\noindent\textit{\textbf{Familiarity with the Topic}:} To understand the participants' experience in performing the tasks, they were asked about their familiarity with  messaging applications and the E2EE feature offered by the apps.

\subsubsection{Main-Study Phase}
During the main-study phase, we presented several code verification challenges to the participants. We also asked them to rate their experience, after finishing each set of the tasks, for a particular app. The six groups of questions (Table \ref{tab:verify}, Appendix \ref{app:tabs}) were presented \textit{randomly} using a 6x6 Latin Square and the questions within each group were also \textit{randomized}. %%verify this randomization
We group the main-study questions into two sets:

\smallskip  \noindent\textit{\textbf{Code Verification Task}:}
For each set of challenges, we asked the participants to follow the instruction on the website to verify the code for a given contact. The list of the verification methods and the steps they need to take to verify the code is presented in Table \ref{tab:verify} (Appendix \ref{app:tabs}). The code verification in our study was performed in one direction (i.e., only the participant compared the code, not the examiner).

\smallskip  \noindent\textit{\textbf{Feedback Questions:}} After completing a set of code verification tasks for a particular application, we asked the participants to rate their experience by answering feedback questions,
%%Say something about SUS
 including System Usability Scale (SUS) questions \cite{brooke1996sus}, and three additional 5-point Likert scale question polled for their comfort, satisfaction, and need for the system. 
% The usability of a system involves several aspects such as effectiveness
%and efficiency of the system, and users’ experience
%6Mann Whitney U test is a non-parametric test suitable for data
%which may not be normally distributed.
%with, and satisfaction of, the system. In general, usability shows
%how much effort and time users should expend to achieve their desired
%objectives and have a satisfactory experience. Based on this
%definition, a simple questionnaire, called System Usability Scale
%(SUS) [17], was designed to measure the usability of an engineered
%system. SUS is a 5-point Likert scale consisting of ten questions,
%each with 5 possible answers (1 represents strong disagreement and
%5 represents strong agreement), covering various aspects of the usability
%of the system, such as the need for support and training, and
%system complexity. SUS score is calculated between 0 and 100,
%and a higher score means better usability.
%At the end of our study, each participant 
 A complete list of the SUS question is included in Appendix \ref{app:sus}. The other three  feedback questions were as follows: 

\begin{itemize}[leftmargin=*]
\item How comfortable are you with this code verification method? 
%(Extremely, Very, Moderately, Slightly, and Not at all comfortable)
\item How satisfied are you with this code verification method? 
%(Extremely, Very, Moderately, Slightly, and Not at all satisfied)
\item Do you want to use this feature in the future? 
%(Everytime, Almost everytime, Occasionally/Sometimes, Almost never, Never)

\end{itemize}

\subsubsection{Post-Study Phase}
In the post-study questions, we asked the participants about their opinion about the code verification task. We also asked them if 
they found any of the code verification methods easier or more difficult than the others.

\section{Analysis and Results}
\label{sec:eval}

%In this section we present and discuss the results and outcomes of our user study.  

\subsection{Pre-Study and Demographics Data} 

We recruited 25 participants from a pool of students in our university. Participation in the study was voluntary and the time allocated to each participant to finish the study was 1 hour. 
%In the user study we recruited 20 participants from the members of our university. 
{There were 54\% males and 46\% females among the  participants in our study. Most of the 
participants were between 18 and 34 years old (34\% 18-24 years, 58\% 25-34 years, and 8\% 35-44 years). 
%The participants were undergraduate and 
%graduate students with diverse educational background including, healthcare, 
%education, engineering, computer science and technology. 
9\% of the participants were high school graduates, 9\% had a college 
degree, 46\% had a Bachelor's degree and 36\% had a Graduate degree.
%Most of them have 
%excellent or good general computer and security skills (
34\% of them declared that they have excellent, and 62\% declared they have good, general computer skills. 17\% had excellent, 50\% had good and 33\% had fair security skills (demographic information is summarized in Table \ref{tab:demog}) in Appendix \ref{app:tabs}.}
About half of the participants said they are extremely to 
moderately aware of the E2EE feature on the apps.
%The participants in our study are young and educated
%students who have declared that they have excellent and good general computer
%and security background, and have used the E2EE feature at least on one of the
%apps prior to the study.

%They said they became aware of the feature through the app notification, news article, or just noticed it while using the app. 
%\remove[MSH]{Table} \ref{tab:familiar} shows the level of 
%familiarity of the participants with different apps and the E2EE feature on them.

Based on the collected data, the most popular apps among the participants 
in our study are WhatsApp, Viber, Telegram, and Signal (these apps are also the 
most popular in the market according to Table \ref{tab:apps}). 
Although the participants declared they have heard about the E2EE feature and used 
it on the apps, it seems they had not performed the code verification task much. 
%This is justifiable since most of these apps started using the feature only recently, 
%and besides, the code verification task is optional in all the apps.
% and maybe the 
%participants have not noticed the importance of code verification and therefore 
%have not performed it much in practice. 

\subsection{Proximity Code Verification Analysis}

%%MSH may remove numeric 
In our study, we tested the code verification task in the proximity setting with QR code 
of Signal (P1-QR), image code 
of Telegram (P2-Image), numeric code of WhatsApp (P3-Numeric). The FAR and FRR for the proximity setting are reported in Figure \ref{fig:labresult}. 

\noindent \textbf{Error Rates:}\textit{ Task P1-QR:} For the QR code presentation, the code verification task is performed  almost automatically. The participants captured the QR code, and after the app verified the code, they reported the success or failure of code verification on the survey form. Since in this model the participant 
is not involved in the task of manual code comparison, we expected the error rates 
(both FRR and FAR) to be near zero. This approach can be considered as the 
baseline of the user study. Indeed, as expected, FRR (instances of rejecting a benign 
session) for the QR code verification is 0\%. However, the FAR (instances of 
accepting an attack session) is 1.34\%. This error happened when one participant 
incorrectly reported that the code verification was successful even though the app deemed it as not matching.

\textit{Task P2-Image:} For the image code verification method, the participants compared the image displayed on his/her screen with the one displayed on the examiner's screen, and verified whether the codes on the two devices match or not. Using this method, the FRR was 0\%, which indicates that none of the benign cases was rejected. The FAR was 2.67\%, only slightly higher than FRR. 
%In the image based verification we did not have control over the number of incorrect characters in the mismatching code. 

%%MSH May remove
\textit{Task P3-Numeric:} For the numeric code verification method, the participant compared the code displayed on his/her screen with the one displayed on the examiner's screen, and verified whether the codes on the two devices match or not. Similar to the image code verification, the FRR was 0\%  while the FAR was 2.67\%.  

%The error appeared in a couple of mismatching instances where only one character was mismatching. 

The Friedman test to compare the FAR among 
multiple proximity code verification methods rendered a Chi-square value of 1.400 and was not statistically significant.\footnote{All results of statistical significance  are reported at a 95\% confidence
level.}

  \noindent \textbf{User Experience and Perception}:  The SUS for the QR code method (66.51) was the  
highest among all the methods in the proximity setting. 
A system with SUS score of below 71 is considered to have an ``OK'' \cite{bangor2009determining}
usability. 
Since the involvement of the users in QR code verification was less compared to the other 
two methods, we expected the users to prefer this method over 
the other two methods. 
To measure the other user perception 
parameters, the participants answered Likert scale questions, rating the comfort, satisfaction, and adoptability from 1 to 5. The answers to this set of  questions are summarized in Figure \ref{fig:labperception}. It seems that,  
after QR code, the numeric
code has the next best user perception ratings.

%%%%%*** support
A Friedman test was conducted to compare SUS, comfort, satisfaction and adoptability, among multiple proximity code verification methods and rendered a Chi-square value of 1.167, 1.214, 1.578, and 4.252 respectively, which were not statistically significant.   
 %
%Further, Wilcoxon signed-rank test, conducted using Bonferroni adjusted alpha levels of 
%0.0167 per test (0.05/3), did not show statistical significance for any of the pairs.

%SUS p=0.497 %%.103 .363 .391
%comfort 0.966 -- 0.617 // Asymp. Sig. (2-tailed) Asymp. Sig. (2-tailed)	.496	.819	.130
%satisfaction 1.411765 --0.494 // Asymp. Sig. (2-tailed)	.168	.592	.163
%need 3.2920.193 -- 0.193 // Asymp. Sig. (2-tailed)	.090	.642	.166

%%%%%*** support

The results of the user study shows low error rates (both FRR and FAR) and
moderate user perception ratings for all the approaches in the proximity
setting, demonstrating that the tested E2EE fared quite well in this setting.

\begin{figure}[t]
  \vspace{-8mm}
    \includegraphics[width=0.48\textwidth]{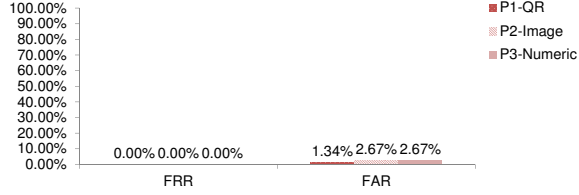}
    \vspace{-4.2cm}
  \caption{{The error rates in performing the code verification task in proximity setting.}}
\label{fig:labresult}
\end{figure}

\begin{figure}[t]
  \vspace{-4mm}
    \includegraphics[width=0.48\textwidth]{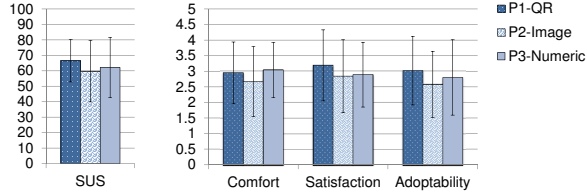}
    \vspace{-4.2cm}
  \caption{{Mean (Std. Dev) of user perception in performing the code verification task in proximity setting.}}
      \vspace{-2mm}
\label{fig:labperception}
\end{figure}

\subsection{Remote Code Verification Analysis}
In the remote setting, we tested audio-based code 
verification of Viber (R1-Audio), image code presentation of Telegram (R2-Image), and numeric 
presentation of WhatsApp (R3-Numeric). The FAR and FRR for the remote setting are reported in Figure \ref{fig:labresult2}.

\noindent \textbf{Error Rates:}
\textit{Task R1-Audio:} In this task, the code was spoken by the examiner 
over a Viber call and the participants compared it with the code displayed on their device. The 
FRR was 0\% in this setting, which shows that none of the valid sessions was rejected by the 
participants. However, the average FAR was on average 26.45\%.
%, which implies that on average, almost one-fourth of the attack cases were not detected by the participants.
{The result also suggests that the error rate increased as the number of mismatching characters decreased as shown in Table \ref{tab:mismatch} (i.e., the success rate of the attack would increase if the attacker has more control over the number of matching and mismatching characters in the code).
%The FAR for 1 mismatching character was 37.27\%. While FAR decreased to 29.32\% when one block was not matching and to 12.75\% when the whole code was not matching. 
Further, we observe that even if the attacker can generate a code that is different from the valid code in only one block, the attacker can compromise about one third of the sessions. }

\textit{Task R2-Image:} 
For the remote image code verification method, in which the image was sent as a text message, the FRR was 18.94\%. This result might lower the usability of the system, since the users would need to repeat the key exchange protocol and the code verification process.
The FAR was 13\%, which indicates that the participants could not detect some of the attack sessions. 
In general, it seems that the users had some difficulty comparing the images. Perhaps they could not memorize and compare the images that appeared on two different apps (i.e., text message app, and Telegram app).
{Since we did not have access to Telegram image generation
source code or tool, we did not have control over the number of mismatching characters. }
%Another reason for the higher FRR and zero FAR could be that users decided to %assume that the images do not match to prevent any possible attack.

\textit{Task R3-Numeric:} For the numeric code verification method, the examiner sent the code in a text message and the participants compared it with the copy of the code that their app generated. 
%Since the users had to switch between the two apps to compare the code, we expected to notice a higher error rates compared to other remote approaches. 
The result shows that the users could not easily compare the numeric codes across the E2EE and SMS app. The average FRR was 22.94\% and the average FAR was 39.69\%. 
Further, the FAR     
increased when the number of mismatching characters decreased, as illustrated in Table \ref{tab:mismatch}. 
%The FAR was 60.25\% for 1 mismatching character, 51.69\% for 1 mismatching block and 7.12\% when the whole code was incorrect.
We saw that the attacker can compromise about half of the sessions, if there is a block of mismatching characters in the code.} %(5 digits, or less than 40 bits)
The  numeric code verification method is deployed by many apps and is one of the major code verification methods used in the remote setting. 
The high error rate in comparing the code shows that such comparison offers low security and possibly low usability.

%%%%%*** support

The Friedman test to compare the FAR among 
multiple remote code verification methods rendered a Chi-square value of 6.306, and was not statistically significant.

%We conducted a non-parametric Friedman test to compare the FAR among 
%multiple remote code verification methods and rendered a Chi-square of 6.306, which was not significant.    %
%Further, Wilcoxon signed-rank test, conducted using Bonferroni adjusted alpha levels of 0.0167
%per test (0.05/3), did not show statistical significance for any
%of the pairs. 
%Thus, we do not observe any statistically significant difference between different remote code comparison methods. 

\begin{figure}[t]
  \vspace{-8mm}
    \includegraphics[width=0.48\textwidth]{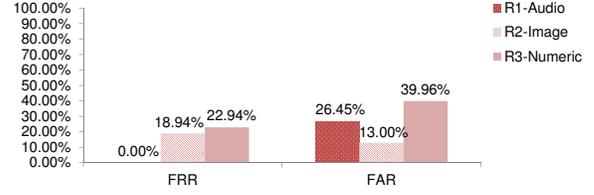}
    \vspace{-4.2cm}
  \caption{{The error rates in performing the code verification task in remote setting.}}
\label{fig:labresult2}
\end{figure}

\begin{figure}[t]
  \vspace{-4mm}
    \includegraphics[width=0.48\textwidth]{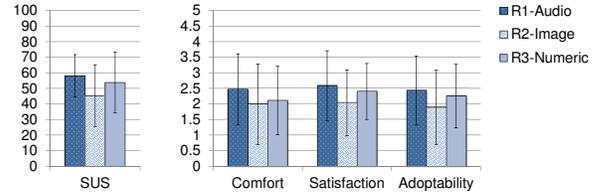}
    \vspace{-4.2cm}
  \caption{{Mean (Std. Dev) of user perception in performing the code verification task in remote setting.}}
      \vspace{-2mm}
\label{fig:labperception2}
\end{figure}

\noindent \textbf{User Experience and Perception:} Viber audio-based code verification received the highest SUS score (58.05) among all the remote code verification methods, followed by WhatsApp numeric code verification (53.78). It seems users did not find Telegram image code verification very usable in the remote setting (SUS=45.53). Systems with SUS scores of below 50 are considered to have ``poor'' usability \cite{bangor2009determining}. Users also seem to rank Viber audio-based code verification the highest with respect to comfort, satisfaction, and adoptability, followed by WhatsApp. Figure \ref{fig:labperception2} summarizes the answers to the user feedback questions.

The Friedman test was conducted to compare SUS, comfort, satisfaction and adoptability score, among multiple remote code verification methods and rendered a Chi-square value of 3.021, 3.381, 4.251, and 6.732 respectively, which were not statistically significant.    
%
%Further, Wilcoxon signed-rank test, conducted using Bonferroni adjusted alpha levels of 0.0167
%per test (0.05/3), did not show statistical significance for any
%of the pairs. 
Thus, based on our study, we do not observe much statistical difference between the different remote code comparison methods in terms of user perception ratings. Most ratings seem to indicate a poor level of usability for all of these methods. 
We recall that comparison between different methods is not the focus of our study. The primary focus is rather to compare between the remote and the proximity settings, which we analyze next.

%SUS 0.329  Asymp. Sig. (2-tailed)	.138	.226	.266
%Comfort 0.123  .235	.406	.061
%Satisfaction 0.191  1.000	.060	.053
%Need 0.065  .628	.161	.053

\begin{comment}
\noindent \textbf{Security and Usability Trade-off:} 
Image code verification with the lowest FAR of 
13\% has the lowest SUS score of 45.35, and lowest comfort, satisfaction and adoptability score, among all the remote code verification approaches, which 
implies higher security and lower usability. On the opposite side, Viber and WhatsApp with higher FAR, seems to have a higher user perception, which shows lower security and higher usability. This result highlights the classical security-usability inherent to a human-centered security system. 
%In Section \ref{sec:discuss}, we discuss how the security can be improved by automating the code verification task while maintaining or improving the usability. 
\end{comment}

\begin{table}[b]
\scriptsize
\centering
\caption{{Effect of number of mismatching character(s) on FAR for remote audio-based and numeric code verification.}}
\vspace{-2mm}
\label{tab:mismatch}
\begin{tabular}{l|r|r|r|}
\cline{2-4}
                                          & \multicolumn{3}{c|}{\textbf{Mismatching Characters}}          \\ \hline
                 \multicolumn{1}{|l|}{{\textbf{Task ID}}}  & \textbf{1 Character} & \textbf{1 Block} & \textbf{Whole Code} \\ \hline \hline
\multicolumn{1}{|l|}{{R1-Audio}}   & 37.27\%              & 29.32\%          & 12.75\%             \\ \hline
\multicolumn{1}{|l|}{{R3-Numeric}} & 60.25\%              & 51.69\%          & 7.12\%              \\ \hline
\end{tabular}
\vspace{-2mm}
\end{table}

%\begin{figure}[h]
%\label{fig:mismatch2}
%\caption{Effect of number of mismatching character on FAR for remote numeric code verification with WhatsApp}
%\includegraphics[width=8cm]{figs/mismatching-whatsapp.pdf}
%\end{figure}

\subsection{Remote vs.\ Proximity Setting}

\noindent \textbf{Error Rates:}
%% here is where most of the statistical will come
%The proximity setting shows zero FRR and a low FAR suggesting that 
%if the proximity setting was to be the primary model, the E2EE apps could have been considered secure from the human errors point of view with ``OK'' usability. 
%However, unlike many other 
%security applications (e.g., device pairing), in practice the proximity setting is not a common model for E2EE apps. For example, two friends may not  
%meet before starting a conversation, which gives the attacker the 
%opportunity to intercept communication while the code has not yet been compared and verified. Installation of the apps also changes the security code. Since friends might not meet to verify the code right after each new installation, this gives the attacker a time window to eavesdrop the communications while the code has not yet been  verified. 
%Given that remote setting is the most common setting in real-life, we refer to the proximity setting as the baseline of the study and compare the error rates and user perception in remote setting to that of proximity setting. 
%Compared to the proximity setting, the remote setting shows a significantly higher FAR. 
%%%%%*** support
%  
The Wilcoxon Signed-Rank test indicates a statistically significant difference between the average FAR and the average FRR of the methods tested in the proximity setting and the methods tested in the remote setting, with the p-value of 0.00 for both FAR and FRR. Thus, in remote setting, the FAR and FRR are
significantly higher than the proximity setting, implying that 
code verification method offers a significantly lower security and a higher chance of rejecting valid sessions. 
The pairwise comparisons also showed statistically significant difference between FAR and FRR in proximity setting and that of remote setting in most cases (detailed statistical analysis can be found in Table \ref{tab:stats} in \ref{app:detailstat}). 
 
%{\color{red} removed FRR and usability}
%Except for the audio-based code verification, which has a 0\% FRR in the remote setting, all the other approaches have significantly higher FRR in remote setting compared to proximity setting. 
%%%%%%*** support
%High FRR {\color{red}may} affect 
%the usability of the system, since valid sessions {\color{red}might} be rejected or restarted. 

%The Wilcoxon Signed-Rank test also showed a statistically significant difference between the average FRR of methods in the proximity settings and methods in the remote setting with the p-value of 0.003. Thus, in remote settings the FRR is 
%significantly higher than FRR in proximity setting. 

 \noindent \textbf{User Experience and Perception:}
Apart from the error rates, responses to the feedback questions (user perception) shows that SUS for all the code verification methods in the remote setting was lower than 
the methods in the proximity setting. Also, users generally were less satisfied and comfortable with the code verification, and were less willing to use the remote  verification methods.

The Wilcoxon Signed-Rank test indicates a significant difference between the average SUS, comfort, satisfaction, and adoptability ratings of methods in the proximity setting and methods in the remote setting, yielding a p-value of 0.005 for the SUS, 0.007 for comfort, 0.004 for satisfaction, and 0.003 for adoptability  (detailed statistical analysis can be found in Table \ref{tab:stats} in \ref{app:detailstat}).
%The Wilcoxon Signed-Rank test did not show statistically significant difference for the adoptability score between the proximity and remote setting. 

%Asymp. Sig. (2-tailed)	.006	.007	.004	.014

%\subsection{Error Rates under Bidirectional Authentication}
\subsection{Bidirectional Authentication}

In the user study, we asked the participants to compare and 
verify the code only in one direction.
In practice, the E2EE applications ask both users involved in the conversation 
to verify each others' code. However, the code verification on the two devices work 
independently, that is, Alice may verify Bob, while Bob does not verify Alice. 
This situation may work in favor of the attacker. If only one party (Alice \textit{or} Bob) 
accepts the attack session, he/she may continue to send messages, files, or photos  
that can be eavesdropped (or tampered) by the attacker. 

Therefore, for the attack to be successful, compromising only one side of the 
connection is enough to intercept several messages (even if we assume that one end-point who does not 
verify the code does not send any messages or does not initiate any call). Hence, in practice, the 
FARs will be higher than the one reported in our study. 
For example, FAR is 40\% in numeric code verification considering the attack in only one direction. Same error reaches 64\% ($=40\%+40\%-40\% \times 40\%$) if the attacker can deceive one of the two parties. 
This result essentially shows that, in practice, in the most common-
use case of E2EE apps (the remote setting), the human errors in comparing the
codes could result in the success of the MITM attack with a high chance.
%, and will make the E2EE security functionality near ineffective.
%
%For example, in the case of numeric code verification with WhatsApp where FAR is about
%40\% in one direction, the error rate increases to 64\% ($=40\%+40\%-40\% \times 40\%$) if the attacker succeeds in any of the two directions.  

\subsection{Post Study Analysis}

At the end of the study, we asked the participants to give us open-ended comments about the study. We also asked them if they would prefer proximity setting over the remote setting. 
The participants commonly agreed that proximity code verification methods in which ``they are sitting side by side'' are easier. Many users found ``QR code to be more user friendly'' than the rest of the methods. 
Some of the participants found the numeric and image code comparison in the remote setting to be ``\textit{frustrating}'' and ``\textit{time consuming}'', as they have to ``switch between windows''.

\section{Discussion and Future Work}
\label{sec:discuss}

%In real-life the apps do not 
%provide the user with explicit information on the security risk associated with 
%neglectful code verification. In fact apps consider the code verification as an optional step 
%that may be taken to avoid MITM attack. 

%\textbf{Summary and Key Insights}
\subsection{Strengths and Limitations of Our Study}

%\noindent \textbf{Lab Study Participants vs. Real-Life Users:}
%\vspace{2mm}
%
We believe that our study has several strengths.
The analysis of the pre-study questionnaire responses shows that the participants in our user study are
above the level of average app users in terms of age, education and computer skills, which suggests that these
participants might have performed better than the average users in detecting the attacks and might have found the
code verification methods more usable (in terms of SUS, satisfaction, comfort,
and adoptability rating) than average users. The participants in our study are young and educated
students who have declared that they have excellent and good general computer
and security background, and have used the E2EE feature at least on one of the
apps prior to the study.  This sampling of user population serves as a sound counter-example to support our
hypothesis that E2EE apps may not provide the promised security and usability,
specially in a remote setting. In practice, these apps are used by a more diverse
group of users who may be less technology-aware, and therefore, may perform worse
than our study participants, thereby granting more opportunity to the attacker to
succeed. Similarly, the average real-world users may find the code verification 
task much less usable compared to our study participants. 

Finally, in the controlled and monitored lab setting, users might be subject to the Hawthorne effect compared to real world settings.  Thus, our work shows that the E2EE apps may provide a lower level of security and usability in 
practice than what was observed in our study.

%\smallskip \noindent \textbf{Primary Task vs.\ Secondary Task:}
%\vspace{2mm}
%
Further, in our study, the participants were involved in one single task of code
verification as the primary task.  At the beginning of the study, we informed
the participants about the security risks of careless code verification and 
asked them to perform the task diligently to avoid the attack as much as
possible.  However, in real-life, the code verification is an optional task in
all the studied apps.  Even those users who verify the codes, may skip through the process (as reported in prior proximity-based device
pairing studied \cite{kuo-fc07}), since their primary task is to establish the 
connection (send a message, or start a call), while verifying
the code is only their secondary optional task.  Therefore, in practice, the
error rates might be higher than the one reported in our studies.
This means that the E2EE apps would be even more vulnerable to attacks and less usable in real-life 
than what our study has found.

%Since, first, the users may not be aware of the consequences of not verifying
%or incorrectly verifying the codes, and second, they may skip the code
%verification task as it is an optional secondary task, or perform it
%neglectfully to pursue  the conversation, which is their primary task.  

%{\smallskip \noindent \textbf{Length of the Code:} 
To achieve
	collision resistance, the current apps aim to use longer codes.
%(e.g., the full public key).  , to make sure that generated codes by the MITM
	%attacker does not match the valid codes. 
However, longer codes are not easy for human users to compare. Therefore,
% To reduce the human errors in comparing the code, 
some apps truncate it to a shorter code. Since we studied the real apps, one limitation of our study 
was that we did
not have control over the length of the code, although these codes is 
what the users are being subject to in real-life. In our study, only Viber code was 160
bits, while other codes were still the same length, 256 bits.
%Also, in Telegram image code verification, we did not have control over the
%number of mismatching characters in the code.  the methods to compare the
%codes were different . 
%%(e.g., 48-digit code as in Viber). 
%one limitation was in telegram you cpuld not control how many bits to change
%antoher is we could not control the bit length of the code

\begin{comment}
\noindent 
\textbf{E2EE Apps on Other Platforms:}
%\vspace{2mm}
%
In this paper, we emphasized on the E2EE apps only on the smartphone platform,
given their popularity. However, many of the end-to-end messaging and VoIP apps
have also deployed applications for desktop or browsers (e.g., Viber Desktop
and WhatsApp Web). The security model of the desktop and browser apps are
usually different from the one offered by the smartphone app. For example, in
WhatsApp Web, the web-client and the device establish a secure connection.  The
smartphone app receives and decrypts the messages, re-encrypts them, and relays
them to the web-client, which decrypts and displays them.  Since the encryption
protocols are different on desktop apps from the E2EE app protocol, we have not
studied the desktop application in this submission. 
These alternative application platforms should be subject to future investigations using 
our study methodology.
\end{comment}

\subsection{Potential Mitigations \& Future Directions}
While our work exposes a fundamental vulnerability of the E2EE apps in the remote setting that arises from the 
human behavior that may be challenging to counter, we sketch two of  
the potential mitigation strategies and directions that may help
curb, if not eliminate, the said vulnerability.

  \noindent \textit{(1) Multi-Windowing:} The post-study open-ended
questions revealed that in the remote setting, many users had difficulty comparing
the codes received through an out of band messaging application due to the
single-tasking interface of the smartphones (i.e, only one application is
displayed on the screen at a given time). Some of the smartphones allow the
users to split the phone's screen into two sections and open one app in each
portion of the screen. Using this feature, for code verification, the user can
open the E2EE app's code verification screen in one portion of the screen and
the out of band messaging app in another portion of the
screen. Since multi-windowing is not available on all smartphones, we have not
evaluated this approach in our study. However, this option would perhaps make the code
verification easier and more robust to errors, since the user can see both
codes at once. This method should be studied as part of future research.  

 \noindent 
\textit{(2) Automated Code Verification:} Human errors in comparing the codes lead to the success of the MITM attacks 
against E2EE apps. 
Apart from informing the users about the 
code verification task and its importance pertaining to the security of the system, 
one natural defense against the attack 
is to reduce the involvement of the human users 
by automating the code verification task itself. Since the threat model assumes trusted 
devices, the code may be verified automatically by the E2EE app by reading directly from
within the out-of-band app. {However, due to privacy issues, this automation may not be securely implemented by allowing the apps to directly access each other's data (e.g., reading the code from the SMS by the E2EE app may create privacy concerns)}.
Further investigation is necessary to explore such a model.
%% NS 2/12: I suggest we look into this permission issue a bit more -- is it allowed in current OSs? I think it is as you said in our meeting. 

%% NS 2/12: commenting out the first sentence.
%In the proximity setting, approaches such as QR codes already exist that can
%effectively eliminate the error rates (as our proximity setting part of the
%study confirms).  However, the remote setting requires a secure  authenticated
%channel, therefore, one cannot simply send the code or QR code over the same
%insecure channel for automatic comparison by the devices.  Since, such code
%transmission scheme is still susceptible to MITM attack. 
{Another possibility for semi-automation is through the use of the clipboard.
	In fact, Signal has most recently (after our study was accomplished)
	implemented a code verification method for the remote setting, in which
	the user copies the received code to the phone's clipboard and asks the
	app to compare and verify the locally generated code with the one
	stored in the clipboard. Therefore, the users do not need to manually
	compare the code. Although this method is a step towards automation,
	further studies are required to evaluate its security and usability.
	For example, one problem with this approach could be that the user
	copies all data currently in the clipboard (which may contain sensitive
	information) and makes it accessible to the E2EE app.}  

\section{Related Work}

%There are numerous studies that evaluated the security and/or usability 
%of code verification  in security applications such as device pairing and key authentication. 
A great number of papers are available that compare different code verification
methods for the device pairing setting (e.g.,  \cite{uzun2007usability,
kumar2009comparative, kainda2009usability}). This line of work mainly targets
the way the code is exchanged over a location-limited out-of-band channel
(e.g., audio vs video), or the way it is compared (e.g., Copy-Confirm vs.
Compare-Confirm). Unlike our study, these studies
assume the exchange of short codes (e.g. 6-digit code) in a \textit{proximity
setting}. 

Other work \cite{hsiao2009study, unicorns} compared different forms of representation
 of cryptographic keys, such as textual and graphical
representations. In our work, we also consider several practical
examples of code presentation, but more importantly, we study the proximity and
remote code exchange, comparison and verification methods in the context of
real-world smartphone apps geared for end-to-end remote security. 
That is, unlike these prior studies, our primary goal is not to compare
across different methods, but rather to compare between the remote and the proximity settings.

Another recent study \cite{acsac-2015} has investigated the security and
usability of the code comparison in end-to-end encrypted VoIP apps in the
presence of data and/or voice MITM attack. Their studies consider short codes
(mainly 2-4 words) that are communicated over the voice channel. Their results
show that users' failure in verifying the code results in accepting on average
about 30\% of the attacked sessions.  While their study covers remote
audio-code verification methods with short words, we consider several long code
presentation and code verification methods that are deployed widely by existing
popular messaging and VoIP applications.  {We did not consider word
representation of codes in our study as these are not deployed by the most
popular apps we focus on in our work, but the analysis of security and
usability of the short word presentation can be found in \cite{acsac-2015}.}

A recent paper by Dechand et al.  \cite{dechandempirical} conducted a
usability study to evaluate performance and usability of the different textual code
representations. In their online study, participants were presented with
several instances of code pairs and were asked to compare the displayed code to
evaluate the performance of the users with respect to the different representations
of the codes (e.g., numeric, hexadecimal and sentence). They concluded that
the use of large words provided the fastest performance, and sentences achieve
the highest attack detection rate. This study does not compare the code
exchange method and code verification process for any specific app. The aim of
this study was to evaluate how easily different representations can be compared from the users' point of view in a \textit{proximity setting}, where two 
codes are shown side by side. Another
limitation is the online study itself that does not capture the exact user interface
of the apps. Our work, in contrast, focuses on the remote setting and its comparison with 
the proximity setting.
%, given that this
%setting is perhaps the most common setting and represents the main purpose of using the
%VoIP and IM apps. 
In addition, our work shows the usability and security of
actual phone apps, not studied before. 
% Finally their study focuses on the benign case rather than the attack case. 

\section{Concluding Remarks}
\label{sec:conclude}

%Internet messaging and calling applications --- representative instances such as
%WhatsApp, Viber, Telegram, and Signal --- are overtaking traditional messaging and
%calling services. These apps have deployed an E2EE feature to provide
%end-to-end security. The E2EE notion critically relies on human users to  verify a security code generated by the protocol to defeat MITM attacks. 

In this paper, we ran a user study to investigate the security and
usability of the human-centered code verification task deployed by a representative class of
E2EE apps.  Our study discloses several security and usability issues stemming
from the human errors in performing the code verification task in a remote
setting. Specifically, we noticed that several code verification methods offer
low security and low usability in the remote setting (much lower than the proximity setting).
%, essentially due to the need to perform cross-app code comparisons. 

Our study design with the security task being the only task at hand and
with well-informed and young participants who performed the security
task diligently and willingly, implies that in real-life situations, the
security offered by these apps could be much lower given that these apps do not
inform the users about the security risks of erroneous code verification, the
real-world users will not be as tech-savvy as our study participants, the
real-world users' primary task will not be security-centric and the real-world users
may rush through the verification process by simply accepting all (including
attacked) sessions. Besides, in real-world, the attacker's success may increase
as she can deceive any of the users involved in a conversation, whereas we
studied the attack only in one direction.  

\begin{comment}
We concluded our work by providing insights for the app designers to improve
the security and usability of the system by automating the code verification
task to lower the impact of the exposed vulnerability.
\end{comment}

%\input{25Result}

%
%
%subsection*{Ackowledgments}
%

\bibliographystyle{ACM-Reference-Format}
\bibliography{all}

\appendix
%% TODO
%%% add survey questions

\section{Appendix}
\label{sec:app-fig}

\subsection{SUS Questionnaire}
\label{app:sus}

\begin{enumerate}
\itemsep0em

\item I think that I would like to use this system frequently.	

\item I found the system unnecessarily complex.

\item I thought the system was easy to use.                      	

\item I think that I would need the support of a technical person to be able to use this system.	

\item I found the various functions in this system were well integrated.

\item I thought there was too much inconsistency in this system.

\item I would imagine that most people would learn to use this system very quickly.			

\item I found the system very cumbersome to use.

\item I felt very confident using the system.

\item I needed to learn a lot of things before I could get going with this system. 

\end{enumerate}

\subsection{Additional Tables and Figures}
\label{app:tabs}

\begin{figure}[h]
\begin{subfigmatrix}{2}
\subfigure[WhatsApp Security Code.]{\includegraphics[width=3.75cm, height=3.75cm]{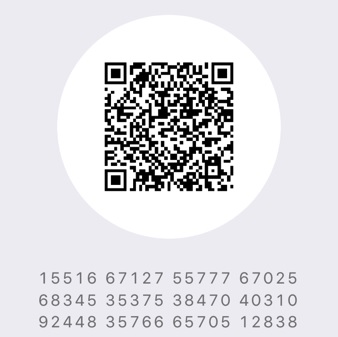} \label{fig:whatsapp}}
\subfigure[Viber Secret Identity Key.]{\includegraphics[width=3.75cm, height=3.75cm]{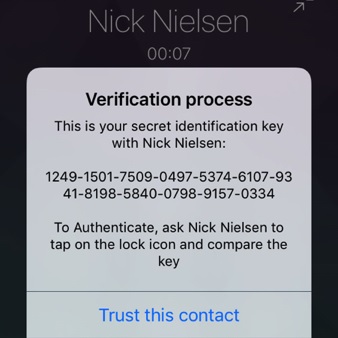}\label{fig:viber}}
\subfigure[Telegram Encryption Code.]{\includegraphics[width=3.75cm, height=3.75cm]{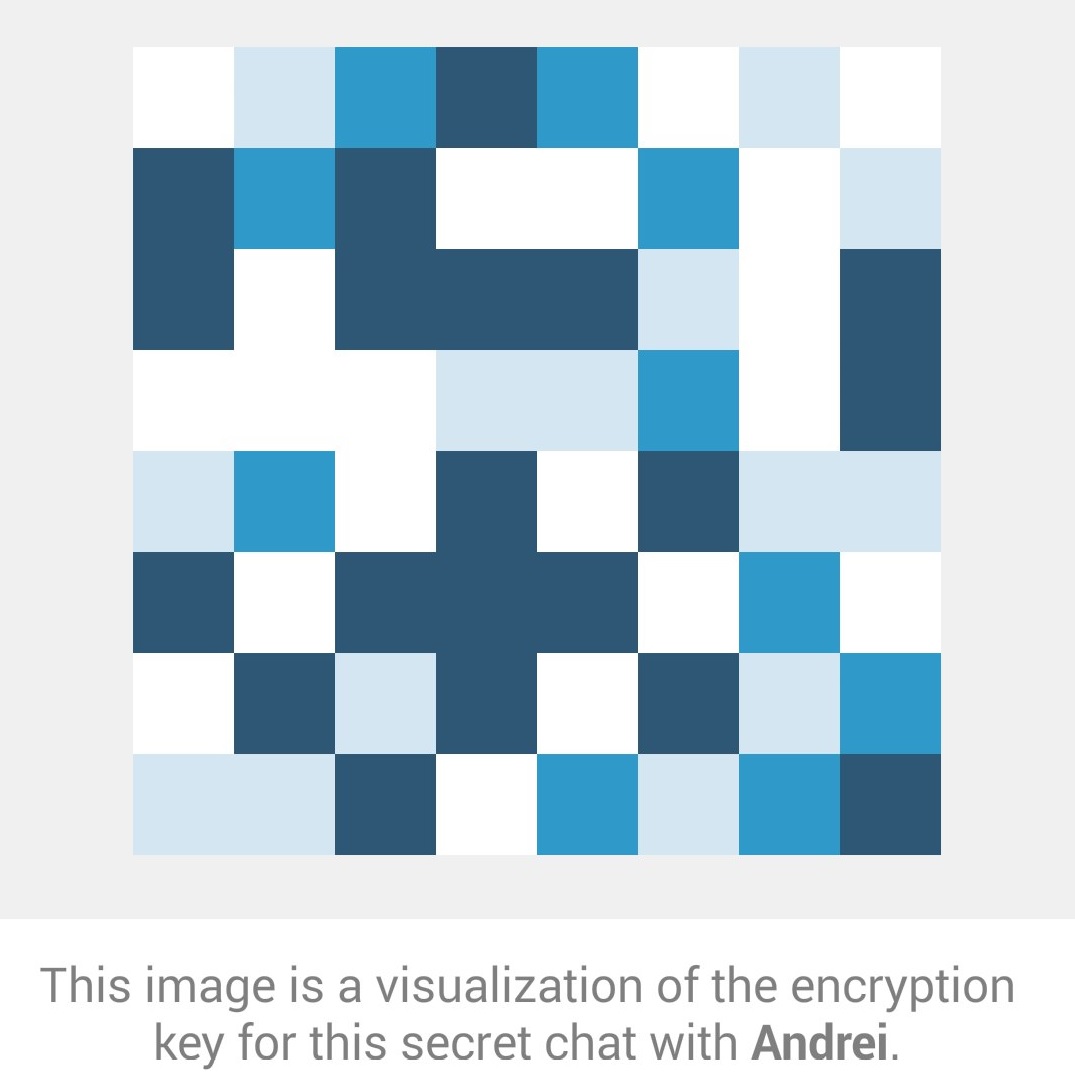}\label{fig:telegram}}
\subfigure[Signal Fingerprint.]
%(earlier versions of the app)
{\includegraphics[width=3.75cm, height=3.75cm]{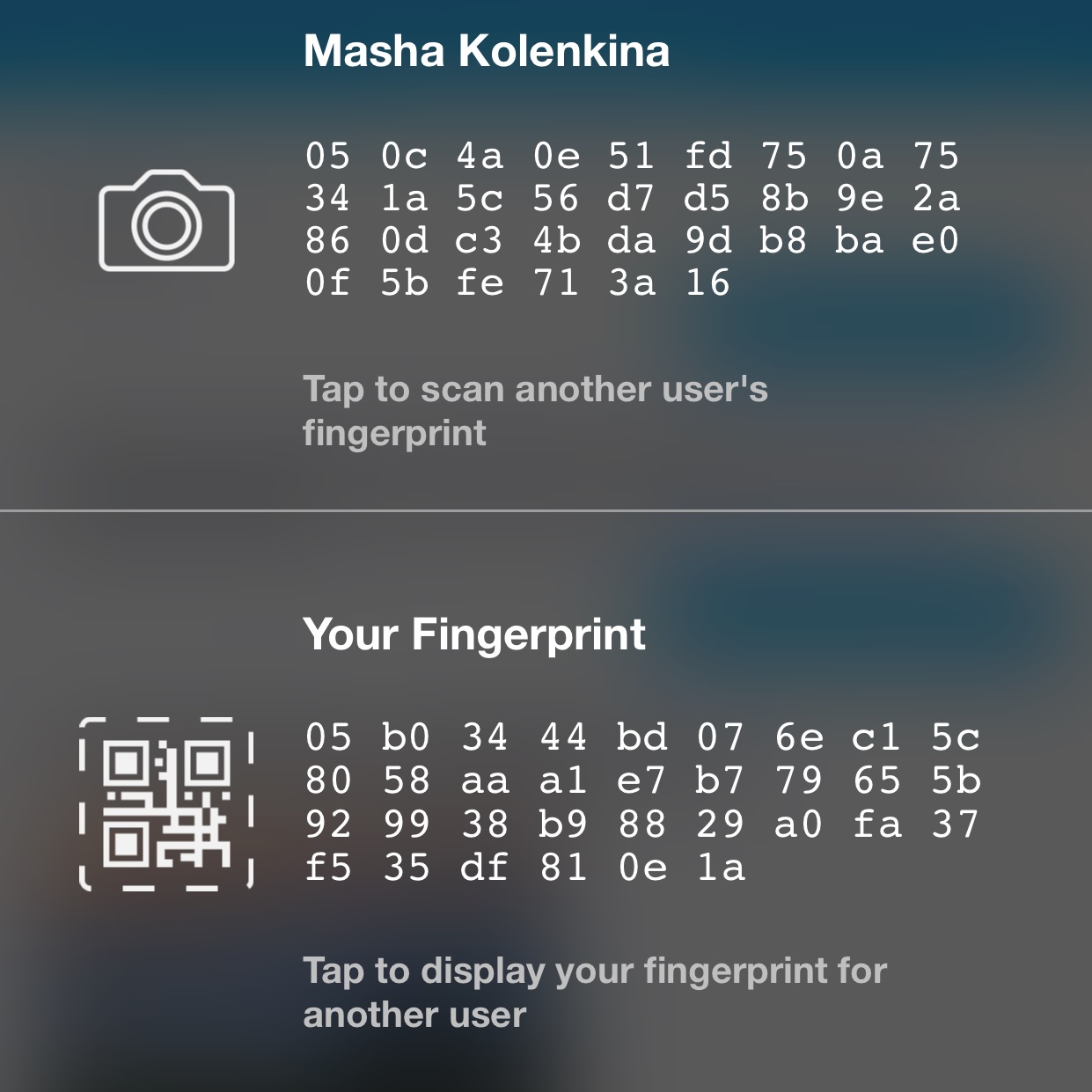}\label{fig:signal}}
%\subfigure[Signal Safety Number  (latest version of the app).]{\includegraphics[width=5cm, height=5cm]{figs/signalSN.jpg}\label{fig:signaln}}
%\subfigure[Silent Circle Short Authenticated String.]{\includegraphics[width=5cm, height=5cm]{figs/silentcircleS.jpg}\label{fig:silentcircle}}
\end{subfigmatrix}
\vspace{-2mm}

\caption{\small{Presentation of the security codes}}
\label{fig:presentation}
\end{figure}

\begin{table*}
  \caption{{The code verification tasks in the user study.}}
      \vspace{-10mm}
%\begin{figure*}[t]
    \includegraphics[width=1\textwidth]{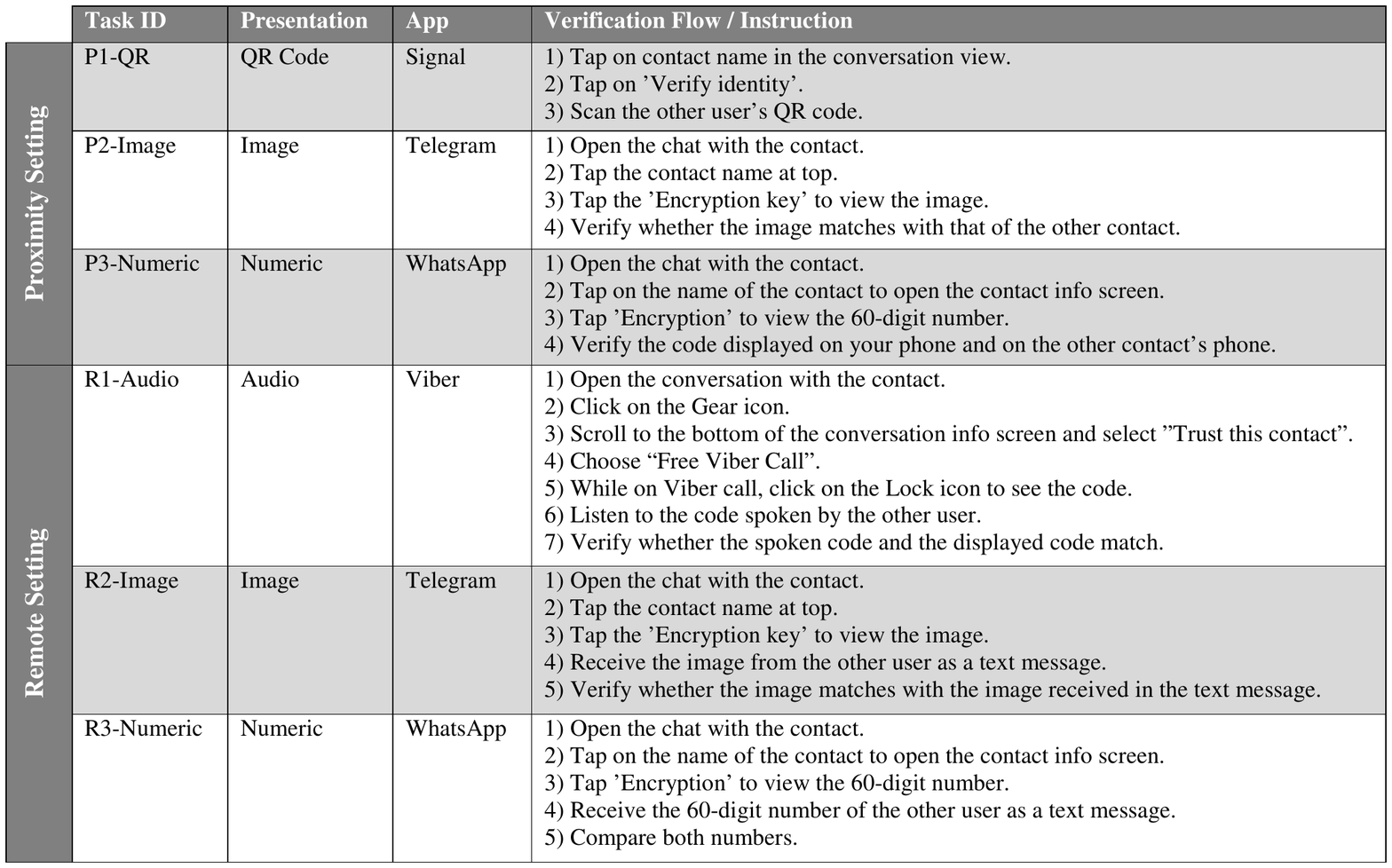}
    \vspace{-4.8cm}
\label{tab:verify}
%\end{figure*}
\end{table*}

\begin{table}[h]
\centering
{\small{
  \centering
  \caption{\small Demographic information of the participants} 
\label{tab:demog}
 
    \begin{tabular}{|rl|r|}

%\hline
%    \multicolumn{2}{|l|}{} & {\bf  Study}\\
%%    \midrule
\hline
    \multicolumn{2}{|l|}{} & {\bf N = 25} \\
%    \midrule
\hline
\hline

%    \multicolumn{2}{l}{Characteristics} & \multicolumn{1}{l}{Overall} \\
    \multicolumn{3}{|l|}{\bf Gender} \\
\hline
          & Male  & 54\% \\
          & Female & 46\% \\
\hline
    \multicolumn{3}{|l|}{\bf Age} \\
\hline
          & 18-24 years & 34\% \\
          & 25-34 years & 58\% \\
          & 35-44 years & 8\% \\
%          & 45-54 years & 0\% \\
%          & 55-64 years & 0\% \\

\hline
    \multicolumn{3}{|l|}{\bf Education} \\
\hline
          & High school graduate or diploma &  9\% \\
          & Some college credit, no degree & 9\% \\
	  & Bachelor's degree  & 46\% \\
          & Graduate degree  & 36\% \\

\hline
    \multicolumn{3}{|l|}{\bf General Computer Skills} \\
\hline
          & Excellent &  34\% \\
          & Good & 62\% \\
	      & Fair  & 4\% \\
          & Poor  & 0\% \\

\hline
\multicolumn{3}{|l|}{\bf General Security Skills} \\
\hline
          & Excellent  & 17\% \\
          & Good & 50\% \\
	      & Fair  & 33\% \\
          & Poor  & 0\% \\
\hline
\multicolumn{3}{|l|}{\bf Awareness about E2EE Feature} \\
\hline
          & Extremely aware  & 17\% \\
          & Moderately aware & 37\% \\
     	  & Somewhat aware  & 16\% \\
          & Slightly aware  & 17\% \\
          & Not at all aware  & 13\% \\
\hline

    \end{tabular}
   \label{table:demog} 
  }}
\end{table}

\subsection{Messaging Apps}
\label{app:apps}

% \usepackage{multirow}
\begin{table*}[]
\scriptsize
\centering
\caption{\small{End-to-End Encrypted Messaging Apps Rating and Reviews}}
\label{tab:apps}
\begin{tabular}{l|l|l|l|l|l|l|l|}
\cline{2-8}
                                            & \textbf{\begin{tabular}[c]{@{}l@{}}Installs on\\ Google Play\end{tabular}} & \textbf{\begin{tabular}[c]{@{}l@{}}Rating on\\ Google Play\end{tabular}} & \textbf{Votes} & \textbf{\begin{tabular}[c]{@{}l@{}}Code\\ Presentation\end{tabular}} & \textbf{\begin{tabular}[c]{@{}l@{}}Code\\ Length\end{tabular}} & \textbf{\begin{tabular}[c]{@{}l@{}}Proximity Code\\ Verification\end{tabular}} & \textbf{\begin{tabular}[c]{@{}l@{}}Remote Code \\ Verification\end{tabular}}                   \\ \hline
\multicolumn{1}{|l|}{\textbf{WhatsApp}}     & \begin{tabular}[c]{@{}l@{}}1,000,000,000 - \\ 5,000,000,000\end{tabular}  & 4.4                                                                     & 45,751,306               & \begin{tabular}[c]{@{}l@{}}QR, \\ Numeric\end{tabular}               & 60 digit                                                       & \begin{tabular}[c]{@{}l@{}}QR code scanning, \\ Manual number compare\end{tabular}       & \begin{tabular}[c]{@{}l@{}}OOB code exchange \\ on messaging apps\end{tabular}                                  \\ \hline
\multicolumn{1}{|l|}{\textbf{Viber}}        & \begin{tabular}[c]{@{}l@{}}500,000,000 - \\ 1,000,000,000\end{tabular}    & 4.3                                                                     & 9,339,793                & Numeric                                                              & 48 digit                                                       & Voice call & Voice call                                                                                                      \\ \hline
\multicolumn{1}{|l|}{\textbf{Telegram}}     & \begin{tabular}[c]{@{}l@{}}100,000,000 - \\ 500,000,000\end{tabular}      & 4.3                                                                     & 2,090,485                & \begin{tabular}[c]{@{}l@{}}Image, \\ Hexadecimal\end{tabular}        & \begin{tabular}[c]{@{}l@{}}64 \\ characters\end{tabular}       & \begin{tabular}[c]{@{}l@{}}Manual number, and \\ image comparison\end{tabular}              & \begin{tabular}[c]{@{}l@{}}Not available directly \\ from the app screen, \\ Any OOB code exchange\end{tabular} \\ \hline
\multicolumn{1}{|l|}{\textbf{Google Duo}}   & \begin{tabular}[c]{@{}l@{}}10,000,000 - \\ 50,000,000\end{tabular}        & 4.3                                                                     & 179,340                  & N/A                                                                  & N/A                                                            & N/A                                                                                         & N/A                                                                                                             \\ \hline
\multicolumn{1}{|l|}{\textbf{Google Allo}}  & \begin{tabular}[c]{@{}l@{}}5,000,000 -\\ 10,000,000\end{tabular}          & 4.2                                                                     & 146,507                  & N/A                                                                  & N/A                                                            & N/A                                                                                         & N/A                                                                                                             \\ \hline
\multicolumn{1}{|l|}{\textbf{Signal}}       & \begin{tabular}[c]{@{}l@{}}1,000,000 - \\ 5,000,000\end{tabular}          & 4.6                                                                     & 86,316                   & \begin{tabular}[c]{@{}l@{}}QR, \\ Numeric\end{tabular}               & 60 digit code                                                  & \begin{tabular}[c]{@{}l@{}}QR Code Scanning, \\ Manual number compare\end{tabular}       & \begin{tabular}[c]{@{}l@{}}Not available directly \\ from the app screen, \\ Any OOB code exchange\end{tabular} \\ \hline
\multicolumn{1}{|l|}{\textbf{Threema}}      & \begin{tabular}[c]{@{}l@{}}1,000,000 - \\ 5,000,000\end{tabular}          & 4.5                                                                     & 44,160                   & \begin{tabular}[c]{@{}l@{}}QR, \\ Hexadecima\end{tabular}            & 32 characters                                                  & \begin{tabular}[c]{@{}l@{}}QR Code Scanning, \\ Manual number compare\end{tabular}       & \begin{tabular}[c]{@{}l@{}}OOB code exchange \\ on messaging apps\end{tabular}                                  \\ \hline
\multicolumn{1}{|l|}{\textbf{Wickr Me}}     & \begin{tabular}[c]{@{}l@{}}1,000,000 - \\ 5,000,000\end{tabular}          & 4.2                                                                     & 9,011                    & N/A                                                                  & N/A                                                            & Video & Video                                                                                                           \\ \hline
\multicolumn{1}{|l|}{\textbf{ChatSecure}}   & \begin{tabular}[c]{@{}l@{}}500,000 - \\ 1,000,000\end{tabular}            & 4                                                                       & 5,211                    & \begin{tabular}[c]{@{}l@{}}QR, \\ Hexadecima\end{tabular}            & 40 characters                                                  & \begin{tabular}[c]{@{}l@{}}QR Code Scanning, \\ Manual number compare\end{tabular}       & \begin{tabular}[c]{@{}l@{}}OOB code exchange \\ on  messaging apps\end{tabular}                                 \\ \hline
\multicolumn{1}{|l|}{\textbf{Silent Phone}} & \begin{tabular}[c]{@{}l@{}}100,000 - \\ 500,000\end{tabular}              & 3.6                                                                     & 1,028                    & Words                                                                & Two words                                                      & Voice call & Voice call                                                                                                      \\ \hline
\end{tabular}
\end{table*}

Table \ref{tab:apps} shows 10 highly popular E2EE apps. The total number of the  installations and the rating of the apps are derived from the Play Store and was last updated on this submission on November 06, 2016. iTunes store does not 
disclose the number of app installations. Although an estimation can be inferred \cite{garg2012inferring}, we believe the current data from Google Play Store serves well to provide information about the popularity of the apps. 

Some recently introduced apps such as Google Duo and Google Allo have not yet deployed any code verification method. For other applications, the code presentations and code verification methods in proximity and remote setting is given in the table. 
Some of the apps such as Telegram and Signal do not offer a feature to directly transfer/exchange the code from the app. Such applications rely on the users to compare the codes through an authenticated out of band channel of their choice. On the other hand, apps such as Viber and Silent phone do not have an explicit way to 
compare the code locally.
%, thus, same approach as the remote setting is used in the proximity setting. 

Considering the popularity of the application and to cover a variety of code presentations and code verification methods, we picked the first four popular apps that offer code verification, namely, WhatsApp, Viber, Telegram, and Signal  for the purpose of this study.

%
%\begin{table}[]
%\centering
%\scriptsize
%\caption{The result ofthe  Friedman test to compare the code verification methods in the remote vs. proximity settings}
%\label{tab:statall}
%\begin{tabular}{l|c|c|}
%\cline{2-3}
%\multicolumn{1}{c|}{}                       & \multicolumn{2}{c|}{\textbf{Friedman Test}} \\ \cline{2-3} 
%\multicolumn{1}{c|}{}                       & \textbf{P Value}    & \textbf{Chi-square}   \\ \hline
%\multicolumn{1}{|l|}{\textbf{FAR}}          & 0.000               & 40.385                \\ \hline
%\multicolumn{1}{|l|}{\textbf{FRR}}          & 0.003               & 18.333                \\ \hline
%\multicolumn{1}{|l|}{\textbf{SUS}}          & 0.000               & 9.579                 \\ \hline
%\multicolumn{1}{|l|}{\textbf{Comfort}}      & 0.150               & 14.090                \\ \hline
%\multicolumn{1}{|l|}{\textbf{Satisfaction}} & 0.034               & 12.072                \\ \hline
%\multicolumn{1}{|l|}{\textbf{Adoptability}} & 0.007               & 15.938                \\ \hline
%\end{tabular}
%\end{table}
\begin{table*}[h]
\scriptsize
\caption{The result of the Friedman and Wilcoxon Signed Rank tests to compare the code verification methods in the remote vs. proximity settings}
\label{tab:stats}
\begin{tabular}{l|c|c|}
\cline{2-3}
\multicolumn{1}{c|}{}                       & \multicolumn{2}{c|}{\textbf{Friedman Test}} \\ \cline{2-3} 
\multicolumn{1}{c|}{}                       & \textbf{p-values}    & \textbf{Chi-square}   \\ \hline
\multicolumn{1}{|l|}{\textbf{FAR}}          & 0.000               & 40.385                \\ \hline
\multicolumn{1}{|l|}{\textbf{FRR}}          & 0.003               & 18.333                \\ \hline
\multicolumn{1}{|l|}{\textbf{SUS}}          & 0.000               & 9.579                 \\ \hline
\multicolumn{1}{|l|}{\textbf{Comfort}}      & 0.150               & 14.090                \\ \hline
\multicolumn{1}{|l|}{\textbf{Satisfaction}} & 0.034               & 12.072                \\ \hline
\multicolumn{1}{|l|}{\textbf{Adoptability}} & 0.007               & 15.938                \\ \hline
\end{tabular}

\begin{tabular}{l||c|c|c|c|c|c|c|c|c|c|c|}
\cline{2-10}
\multicolumn{1}{c|}{}                       & \multicolumn{9}{c|}{\textbf{p-values of Wilcoxon Signed Rank Test}}                                                                                                                                                                                                                                                                                                                                                                                                                                                                                                                                                                                                                                                               \\ \cline{2-10} 
\multicolumn{1}{c|}{}                       & \textbf{\begin{tabular}[c]{@{}c@{}}R1-Audio\\ vs.\\ P1-QR\end{tabular}} & \textbf{\begin{tabular}[c]{@{}c@{}}R1-Audio\\ vs.\\ P2-Image\end{tabular}} & \textbf{\begin{tabular}[c]{@{}c@{}}R1-Audio\\ vs.\\ P3-Numeric\end{tabular}} & \textbf{\begin{tabular}[c]{@{}c@{}}R2-Image\\ vs.\\ P1-Q1\end{tabular}} & \textbf{\begin{tabular}[c]{@{}c@{}}R2-Image\\ vs.\\ P2-Image\end{tabular}} & \textbf{\begin{tabular}[c]{@{}c@{}}R2-Image\\ vs.\\ P3-Numeric\end{tabular}} & \textbf{\begin{tabular}[c]{@{}c@{}}R3-Numeric\\ vs.\\ P1-QR\end{tabular}} & \textbf{\begin{tabular}[c]{@{}c@{}}R3-Numeric\\ vs.\\ P2-Image\end{tabular}} & \textbf{\begin{tabular}[c]{@{}c@{}}R3-Numeric\\ vs.\\ P3-Numeric\end{tabular}} \\ \hline
\multicolumn{1}{|l|}{\textbf{FAR}}          & 0.007                                                                   & 0.011                                                                      & 0.010                                                                        & 0.063                                                                   & 0.081                                                                      & 0.066                                                                        & 0.001                                                                     & 0.001                                                                        & 0.001                                                                          \\ \hline
\multicolumn{1}{|l|}{\textbf{FRR}}          & 0.000                                                                   & 0.010                                                                      & 0.010                                                                        & 0.046                                                                   & 0.046                                                                      & 0.046                                                                        & 0.025                                                                     & 0.025                                                                        & 0.025                                                                          \\ \hline
\multicolumn{1}{|l|}{\textbf{SUS}}          & 0.085                                                                   & 0.614                                                                      & 0.637                                                                        & 0.016                                                                   & 0.099                                                                      & 0.033                                                                        & 0.008                                                                     & 0.551                                                                        & 0.150                                                                          \\ \hline
\multicolumn{1}{|l|}{\textbf{Comfort}}      & 0.380                                                                   & 1.000                                                                      & 0.177                                                                        & 0.029                                                                   & 0.048                                                                      & 0.093                                                                        & 0.013                                                                     & 0.0191                                                                       & 0.018                                                                          \\ \hline
\multicolumn{1}{|l|}{\textbf{Satisfaction}} & 0.133                                                                   & 0.874                                                                      & 0.156                                                                        & 0.007                                                                   & 0.027                                                                      & 0.008                                                                        & 0.070                                                                     & 0.902                                                                        & 0.233                                                                          \\ \hline
\multicolumn{1}{|l|}{\textbf{Adoptability}} & 0.068                                                                   & 1.000                                                                      & 0.226                                                                        & 0.006                                                                   & 0.046                                                                      & 0.022                                                                        & 0.017                                                                     & 0.565                                                                        & 0.110                                                                          \\ \hline
\end{tabular}
\end{table*}

\subsection{Statistical Analysis of Remote vs.\ Proximity Setting}
\label{app:detailstat}

The non-parametric Friedman test followed by the Wilcoxon signed-rank test  with Bonferroni correction was conducted to compare the error rates and the user perception among different code verification methods in remote and proximity setting. The results are summarized in Table \ref{tab:stats}.

\end{document}